\documentclass{svjour3}  

\usepackage{authblk}
\usepackage{float}
\usepackage[hidelinks]{hyperref}
\usepackage[sort&compress]{natbib}
\setcitestyle{authoryear}
\usepackage{caption}
\usepackage{graphicx}%
\usepackage{multirow}%
\usepackage{amsmath,amssymb,amsfonts}%
\usepackage{mathrsfs}%
\usepackage[title]{appendix}%
\usepackage{xcolor}%
\usepackage{textcomp}%
\usepackage{manyfoot}%
\usepackage{booktabs}%
\usepackage{algorithm}%
\usepackage{algorithmicx}%
\usepackage{algpseudocode}%
\usepackage{listings}%
\usepackage{algorithm}%
\usepackage{lstcustom}
\usepackage{hyperref}
\usepackage{todonotes}
\usepackage[most]{tcolorbox}
\setuptodonotes{inline}

\makeatletter
\def\makeheadbox{%
  \hbox to0pt{%
    \vbox{\baselineskip=10dd
      \hrule
      \hbox to\hsize{%
        \vrule\kern3pt
        \vbox{\kern3pt
          \hbox{\bfseries Empirical Software Engineering}%
          \hbox{(accepted manuscript)}%
          \kern3pt}%
        \hfil\kern3pt\vrule}%
      \hrule}%
    \hss}}
\makeatother

\date{Accepted: 24/03/2025}

\title{Information-Theoretic Detection of Unusual Source Code Changes}

\author{Adriano~Torres \and Sebastian~Baltes \and Christoph~Treude \and Markus~Wagner}

\institute{%
  Adriano~Torres \at
  University of Adelaide, Australia
  \and
  Sebastian~Baltes \at
  University of Bayreuth, Germany\\
  \email{sebastian.baltes@uni-bayreuth.de} (corresponding author)
  \and
  Christoph~Treude \at
  Singapore Management University, Singapore
  \and
  Markus~Wagner \at
  Monash University, Melbourne, Australia
}

\begin{document}

\maketitle

\abstract{The code base of software projects evolves essentially through inserting and removing information to and from the source code. We can measure this evolution via the elements of information---tokens, words, nodes---of the respective representation of the code. In this work, we approach the measurement of the information content of the source code of open-source projects from an information-theoretic standpoint. Our focus is on the entropy of two fundamental representations of code: tokens and abstract syntax tree nodes, from which we derive definitions of textual and structural entropy. We proceed with an empirical assessment where we evaluate the evolution patterns of the entropy of 95 actively maintained open source projects. We calculate the statistical relationships between our derived entropy metrics and classic methods of measuring code complexity and learn that entropy may capture different dimensions of complexity than classic metrics. Finally, we conduct entropy-based anomaly detection of unusual changes to demonstrate that our approach may effectively recognise unusual source code change events with over 60\% precision, and lay the groundwork for improvements to information-theoretic measurement of source code evolution, thus paving the way for a new approach to statically gauging program complexity throughout its development.}

\keywords{Information theory, entropy, software engineering, source code analysis}

\section{Introduction} \label{introduction}
Throughout the evolution of a software project, it is both natural and expected that applications be able to change and adapt to suit its users if it is to remain useful ~\citep{Lehman1980Laws}. As the development life cycle unfolds, maintaining a project's code base entails activities such as implementation of new behaviour, removal of features, bug corrections, or architectural and performance enhancements. 

Given that change is inherent in the life cycle of applications that produce value for sufficiently long periods of time and that most changes to software occur during the maintenance period, usually its longest and most costly phase~\citep{LEHMAN1984ProgramEvolution}, it becomes important to create systems that help us assess, manage, and adapt to constantly evolving software. In this work, we introduce an information-theoretic approach to measuring the effect of the ongoing process of software development on its source code.

\subsection{Software evolution and its effects} \label{evolution_effects}
Developers write software in high-level programming languages primarily by positioning tokens in specific positions. The continual change that source code is subjected to during maintenance means that its vocabulary is variable over time. This can manifest itself from a \textit{textual} perspective, where tokens may change in frequency over time, as well as \textit{structurally}, where Abstract Syntax Trees (AST) and Control Flow Graphs (CFG)~\citep{aho86dragon} nodes change in frequency.

Each change event is a potential source of bugs, unnecessary complexity, or unpredictable behaviour, making complexity management a core aspect of software construction~\citep{McConnell:2004:CCS:1096143}. Uncontrolled complexity is generally a byproduct of external demands or inadequate design and management and is a common source of technical debt~\citep{mcconnel2018techdebt}. Researchers have proposed metrics to act as proxies for source code complexity, deriving both from raw source code, like line, token and method count, as well as from intermediate representations, such as McCabe's cyclomacic complexity which is derived from the CFG~\citep{mccabe1976complexity}. Halstead proposed statistical measurements based on the frequencies of operators and operands extracted from the source code as a text stream, and calculated measures of code length, difficulty to write and review, and effort taken to write the code~\citep{halstead1977elements}.

During the development process, engineers work together to maintain the project. It thus becomes important that the different stakeholders involved in the process be kept aware of the changes that are happening to the code, as well as their consequences~\citep{Dourish1992Awareness}. An increase in the number of contributors presents certain challenges to the maintenance process and can be correlated with loss of productivity~\citep{Blackburn1997Productivity}. Furthermore, more code changes are going to be submitted simultaneously (and, in general, more events will happen in the repository), which means that it will be hard for maintainers to be aware of all such events and react accordingly to them. Surveys of developer behaviour and productivity indicate that maintainers could benefit from tools that automate the classification and ranking of events in the project's repository, as well as from notifications regarding \textit{unusual events}~\citep{Treude2015DevActivity}. 

\begin{figure}
    \centering
    \includegraphics[width=\textwidth]{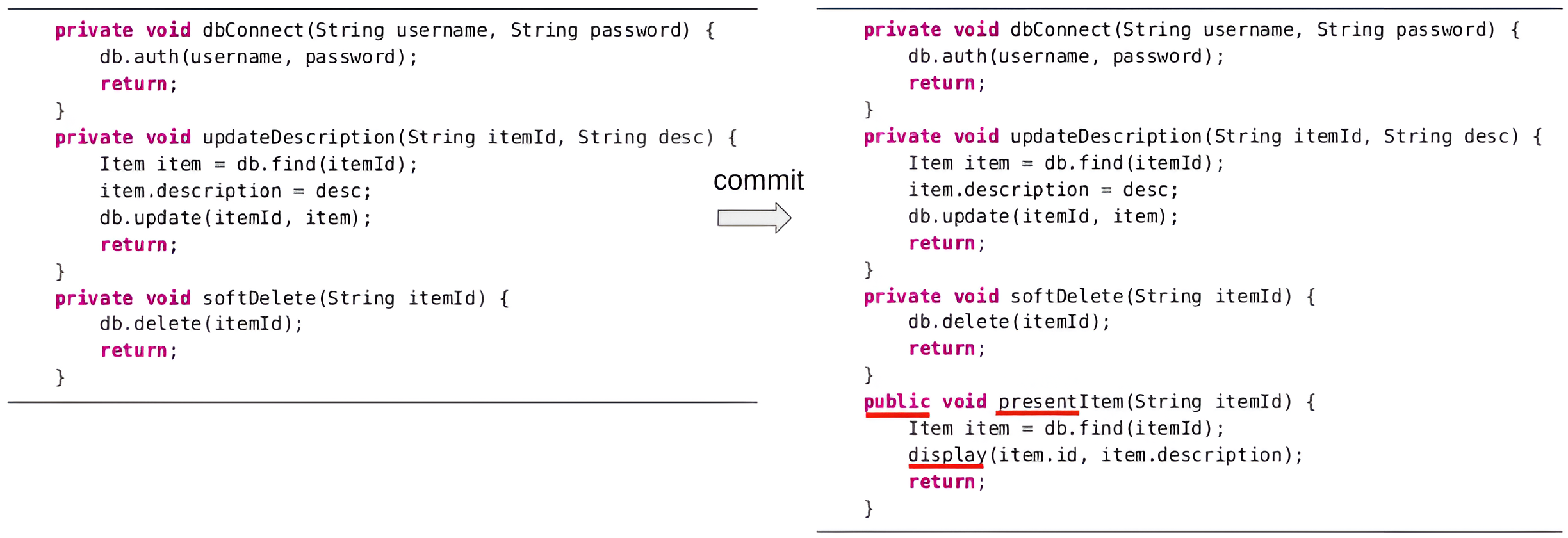}
    \caption{Introduction of a new operation on a data access class}
    \label{fig:expected-change}
\end{figure}

To illustrate the notion of different expectations about source code changes, let us examine an example. In Figure \ref{fig:expected-change}, we extend the behaviour of a data access class by adding a method. The newly added \texttt{hardDelete} method is very similar, both in structure and text, to \texttt{softDelete}. The proposed operation seems adequate to the context of this file.

\begin{figure}
    \centering
    \includegraphics[width=\textwidth]{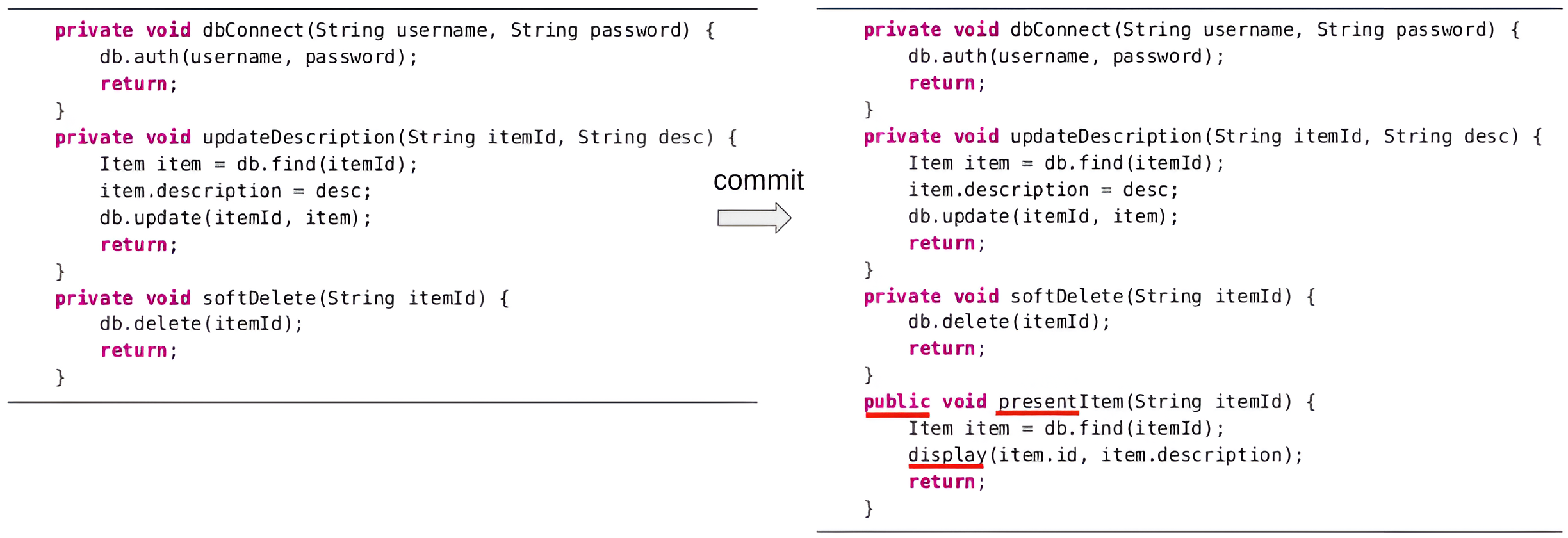}
    \caption{Introduction of a new concern}
    \label{fig:unexpected-change}
\end{figure}

Let us consider a case when the proposed change is the one in Figure \ref{fig:unexpected-change} instead. In this case, the new \texttt{presentItem} method has access permissions that differ from its siblings, and makes a delegate call that does not invoke \texttt{db}. Furthermore, considering the context of a data access class, one could argue that the words \texttt{present} and \texttt{display} are less expected in the context than \texttt{hard} and \texttt{remove}. Regardless of both changes being syntactically valid and producing behaviour that may be correct, the second change may be deemed more unusual or \textit{surprising} than the first. In the rest of this work, we investigate how information theory can help us quantify our expectations about these events and derive measures of source code complexity.

\subsection{Information theory and entropy} \label{motivation_information_theory}

Claude Shannon originally proposed information theory to address problems of communication over noisy channels, as well as to measure the amount of information conveyed by the different messages that are exchanged over a communication channel~\citep{shannon1948theory}. Its central concept is the Shannon entropy and it defines a way of measuring how much information is produced by an event. Given a set of events $E$ and their respective probabilities, the Shannon entropy is defined as \footnote{One may choose any logarithmic base. We consider base 2 in this work, which results in the bit being the of measure of entropy.}:

\begin{equation} \label{shannon:entropy}
    H(E) = -\sum\limits_{e \in E} p(e) * \log_2 {p(e)},
\end{equation}

The way this definition naturally quantifies expectations follows from the fact that events with lower probabilities produce higher entropy. Considering the simplest case of two events with equal probability, as in a fair coin toss, we obtain a total entropy of 1 bit, and each of the possible events correspond to half a bit. If we remove uncertainty from the system by skewing probabilities to 70\% and 30\%, our resulting total entropy will be about 0.881, with the most likely event contributing 0.36022 bits, whereas the less likely event represents 0.52078 bits. In other words, the less probable events, or the \textit{less expected ones}, are the ones that produce more information.

From the standpoint of information theory, we can see a program's source code as a discrete \textit{information source}, to which there exist multiple receivers. Not only is it consumed by engineers working on the project's development, but it also consumed by software itself, since high level, natural language-like source code is the input in front ends of programs that generate and execute code (i.e., compilers and interpreters)~\citep{aho86dragon}. It thus follows that, as a project is maintained, the capacity of the of the channel that represents the program varies over time. The code base's total information content at any point in time can be calculated by the Shannon entropy:

\begin{equation} \label{source:entropy}
H(X)= \sum\limits_{s \in S} p(s)* \log_2 {p(s)},
\end{equation}

for each symbol $s$ in the symbol set $S$. At any point in time, the set $S$ is defined by the unique elements that compose the alphabet of the chosen representation of source code, as well as their frequencies. When we see source code from its highest level representation (a sequence of words), $S$ is the set of unique words $W$ present in the code. If we instead attain ourselves to intermediate representations, the set $S$ may be, for example, the set of nodes of the program's Control Flow Graph.

Our definition of \textit{textual} entropy is the one suggested in the literature~\citep{berlinger1980complexity, cook1993information}:

\begin{equation} \label{h_token}
H_{TOKEN} = -\sum\limits_{word \in W} p(word) * \log_2 {p(word})
\end{equation}

Analogously, we introduce two definitions of \textit{structural} entropy obtained from the edges of the respective AST of a program.

\begin{equation} \label{h_astedge}
H_{AST\_EDGE} = -\sum\limits_{edge \in E} p(edge) * \log_2 {p(edge})
\end{equation}

\subsection{Contributions} \label{contributions}
In this paper we the following contributions:
\begin{itemize}
    \item A systematic literature review of the status quo of applications of information theory to software engineering;
    \item An analysis of the evolution of the entropy of the source code files for two different representations of source code: as AST elements and as token streams;
    \item An empirical assessment of the evolution of source code entropy conducted on 95 actively maintained Java projects hosted on GitHub. We also measure the impact of different ways of grouping the tokens that compose the code, as well as the correlations between entropy metrics and widely accepted metrics of software complexity. 
    \item A data-oriented approach, using the information-theoretic measurements produced by our definitions, to analyse how Lehman's laws of software evolution~\citep{Lehman1980Laws} apply to the current landscape of software development.
    \item An entropy-based method for detecting and classifying unusual changes to a project's information content.
\end{itemize}

In this paper, we provide significant extensions to our previous work published in the 2nd Intl. Workshop on NL-based Software Engineering~\citep{nlbse2023entropy}:
\begin{itemize}
    \item We almost quadrupled the number of projects analysed, from 25 projects to 95, thus collecting more data to compute gradients, calculate correlations and detect trends;
    \item Upon further analysing the data generated in our previous work, we opted to change our outlier detection approach so as to increase its precision towards values that developers could consider acceptable~\citep{Christakis2016DevelopersNeeds};
    \item We perform quantitative assessments, using the historical data we collected, of postulates that have been posited to govern the evolution of software code bases,~\citep{Lehman1979Understanding};
    \item We categorise and analyse the events that can be detected as unusual variations of source code entropy. We assess the precision of the detection and discuss approaches that can further improve the detection.
\end{itemize}

\subsection{Research Questions} \label{research_questions}
It is generally accepted that the complexity of applications that remain useful increases over time, and that quantitative measurements of such evolution for different projects may provide practitioners with insights into strategies for better complexity management, which can have considerable impacts on the project cost~\citep{LEHMAN1984ProgramEvolution}. This leads us to our first research question.

\begin{tcolorbox}[colback=blue!5!white,colframe=black!75!black,title=Research Question 1 (RQ 1): What are the trends of evolution of source code entropy of actively maintained projects?]
\begin{itemize}
    \item \textbf{Research Question 1.1 (RQ 1.1)}: What are the different patterns in the evolution of source code entropy?
    \item \textbf{Research Question 1.2 (RQ 1.2)}: What are the similarities and differences in the trends of evolution of distinct projects?
\end{itemize}
\end{tcolorbox}

Given the inherent complexity involved in the construction of large-scale software~\citep{McConnell:2004:CCS:1096143}, it is reasonable to assume that having more methods to quantify such complexity can assist in its management, especially if these methods uncover different dimensions of complexity than the ones captured by currently used metrics. This is the subject of our second question.

\begin{tcolorbox}[colback=blue!5!white,colframe=black!75!black,title=Research Question 2 (RQ 2):]
How do information theory-based metrics of source code entropy correlate to standard metrics of software complexity?
\end{tcolorbox}

Lehman's laws of software evolution, established in the 1980s, are frequently referenced in the literature of program evolution, and they make intuitive sense~\citep{Lehman1979Understanding, Lehman1980Laws}. However, the context in which they were derived was remarkably different from today's landscape of software development. Not only were they stated at a time when not much code was publicly available for enterprise application, but compute resources to statistically analyse code bases were not as abundant as they are today. This limits the extent to which such postulates could be empirically validated. Our third research question aims to leverage source code and compute available today to quantitatively assess Lehman's laws.

\begin{tcolorbox}[colback=blue!5!white,colframe=black!75!black,title=Research Question 3 (RQ 3):]
To what extent do information-theoretic measurements of source code evolution via entropy agree with Lehman's laws of program evolution?
\end{tcolorbox}

Being aware of and reacting properly to important events in a repository is one of the most important roles of its maintainers. Provided that capturing and notifying unusual events can help them manage the amount of information generated in a repository~\citep{Treude2015DevActivity}, we can leverage the innate ability of information theory to quantify expectations and apply it to detect and classify unusual change events, which is what we set out to answer in our last research question.

\begin{tcolorbox}[colback=blue!5!white,colframe=black!75!black,title=Research Question 4 (RQ 4):]
What kind of events can we detect by applying entropy-based anomaly detection?
\end{tcolorbox}

\section{Related Work} \label{literature_review}
To form a picture of the state of the art of applications of information theory to software engineering measurement, and more specifically to source code measurements, we conducted a survey of the literature published on the topic from Shannon's~\citep{shannon1948theory} original publication in 1948 to 2023. In this section we present the results of applying Kitchenham's~\citep{Kitchen2009LitReview} protocol for conducting systematic literature reviews in software engineering to evaluate applications of information theory to software engineering.

In our survey, we set out to attain satisfactory answers to the following review questions:
\begin{itemize}
    \item What is the state of maturity of information-theoretic measurements of source code?
    \item To what extent have the current have been verified against empirical evidence?
    \item What are the research opportunities in the field?
\end{itemize}

\subsection{Procedures} \label{review_procedures}
Starting from the \textit{dblp} computer science bibliography repository the starting point of our review, we downloaded a snapshot of all titles published until 28 July 2023 and performed successive regular expression searches with the \texttt{grep} utility.
Starting with a data set of 10,059,878 titles, we applied eight separate queries that resulted in 9031 titles, which were then traversed manually to filter out unrelated titles. Since this process filtered out almost 98\% of the results, we then read the abstracts of the remaining 216 publications and applied another round of manual filtering which consisted in reading the introduction and methods sections of the articles and only keeping the ones who were applying information theory to measuring either source code itself or structures that represent software systems (e.g., class diagrams, design and specification tables, module graphs). This left us with 98 potentially related papers. Table \ref{review-filter} summarises the results of each initial query query. The list of titles resulting from each query, as well as the outcome of manually visiting each of the 98 papers is provided in the supplementary material \cite{supp_material}.

\begin{table}
\centering
\begin{tabular}{cccc}
 \toprule
 Pattern & Titles & Potentially relevant & Relevant\\ \midrule
 information theor* & 3206 & 35 & 15\\ \midrule
 information-theor* & 2100 & 34 & 14\\ \midrule
 program AND complexity & 869 & 31 & 16\\ \midrule
 software AND complexity & 484 & 73 & 19\\ \midrule
 code AND complexity & 1912 & 15 & 8\\ \midrule
 program AND entropy & 76 & 4 & 3\\ \midrule
 software AND entropy & 97 & 24 & 23\\ \midrule
 code AND entropy & 287 & 0 & 0\\ \midrule
 Total & 9031 & 216 & 98\\ \bottomrule
\end{tabular}
\caption{Results of search for patterns in the dblp database.}
\label{review-filter}
\end{table}

To remove the remaining unrelated works, we read all of the 98 resulting publications and identified that 16 of them were pertinent to source code measurement through information theory. For each paper, we created a summary of the work's strengths and weaknesses, assessed whether they validated their claims empirically, and finally categorised the papers between theoretical, empirical and perspective. Short papers consisting of 5 or less pages were disregarded. We applied one level of snowballing to each of these papers, which uncovered 3 more references. Although none of them were indexed by dblp, we only discarded two of them. The paper that we kept was published by Khoshgoftaar et al.~\citep{Khoshgoftaar1994ApplicationsOI}, whose work was, according to our searches, the most relevant set of papers in the field. Khoshgoftaar's paper also presented a survey of the hitherto status quo (1994), and was influential to our own in several ways.

\subsection{Review Summary} \label{review_summary}
In this section we present a chronological description of the most relevant publications on information-theoretic measurements of source code, and we also investigate how our survey process addresses the questions that guided our literature review.

\subsubsection{Overview} \label{review_overview}
We can divide the history of research in the field in two main periods:
\begin{itemize}
    \item \textbf{1970-2009}: Theoretical explorations and proposals on how to apply information theory to software engineering measurement, usually with restricted empirical validation. During this period, devising the metrics themselves was the most pressing concern. Importantly, some of the hypotheses and assumptions about the coding process, such as that it is a stochastic process~\citep{chen1978complexity}, or that all operators were equally likely~\citep{halstead1977elements}, were taken as approximations, for computing resources to these was not readily available;
    \item \textbf{2009-2023}: Following Hassan's ~\citep{hassan2009predicting} attempts to predict fault insertion by via measuring entropy of metadata about the code (e.g., probabilities of files being changed, number of lines added and/or removed), researchers started to investigate how such metrics impact the evolution of software, as well as how machine learning models could assist in predicting fault insertion~\citep{Kaur2017EntropyMl}. Attempts to measure the impact of refactoring efforts arrived at different results, with some authors claiming that it stabilises entropy~\citep{canfora2014changes}, whereas others concluded that~\citep{keenan2022investigation} it tends to increase entropy. In this period, a single study that tried to measure the entropy of the representations of the source code was published, but its results were not validated in real-world production software~\citep{keenan2022investigation}. As mentioned previously, the papers from this period do not attempt to make measurements about the source code itself, instead focussing on how attributes of the code change over time.
\end{itemize}

Although the literature on attempts to measure software complexity through information theory dates back to 1970s~\citep{hellerman1972lookup}, most of its recent applications do not consider source code directly. Hellerman's~\citep{hellerman1972lookup} definition of entropy based on tables that map inputs to outputs are the first recorded explorations of entropy in software engineering. In the years that followed, most research still focused on metrics regarding descriptions of software systems other than its source code, such as specification sheets~\citep{Paulson1992spec} and object-predicate tables~\citep{Mohanty1981EntropyDesign}.

The earliest proposals to define entropy-based metrics of source code date back to 1980, when Berlinger assumed a programming environment that had the ability to keep track of frequencies of operators, identifiers and labels~\citep{berlinger1980complexity}, and proposed that the entropy of the tokens present in the source code could be used as a proxy for programme complexity. Though they claim that a such a mathematically sound definition of information content was expected to correlate well with difficulty to understand a code base, no empirical assessments were provided. 

Chen proposed to represent programs as graphs containing three possible different types of node: sequence, if-else and loop, whose frequencies would be used to calculated program entropy~\citep{chen1978complexity}, whilst assuming programming to be a stochastic process that produces all these control structures with the same probability. 

Davis~\citep{davi88complexity} proposed dividing programs into sequences of contiguous program statements called chunks, which do not transfer control to statements outside of itself (a notion that resembles a control flow graph node~\citep{aho86dragon}), and calculating entropy based on three varying aspects of a chunk: its connection to other chunks, the content of the chunk, and its size. 

In the 1990s, Khoshgoftaar et al. produced a survey of the status quo of graph entropy and its applications to software engineering measurement~\citep{khoshgoftaar1998information}, and followed up with explorations of information-theoretic approaches to quantifying coupling and cohesion of idealised graph abstractions of software~\citep{Khoshgoftaar2001Coupling}. As we shall see in \autoref{review_conclusions}, the overall state of research since their survey has not evolved considerably. 

Recent publications in the field have followed a slightly different route, by focusing primarily on metadata regarding the code, and not the source code itself. Hassan introduced the concept of change entropy~\citep{hassan2009predicting}, wherein entropy is calculated based on the probabilities of files being changed. Canfora et al.~\citep{canfora2014changes} built on the idea by slicing program histories' into blocks of 500 commits and estimating how entropy evolves for 4 open source projects, and concluded that refactoring efforts tend to decrease entropy, as well as that files touched by many developers tend to exhibit higher change entropy. However, researchers exploring change entropy and using 100-commit time slices arrived at opposite conclusions~\citep{keenan2022investigation}. We believe that not considering evolution on a commit-by-commit basis may yield inaccurate and unpredictable results, as virtually every modification to a file should affect its overall information content (provided precise enough measurements). Other researchers followed up with attempts to predict the entropy of code changes~\citep{Chaturvedi2014Complexity}, as well as building machine learning models to predict bug insertion based on change entropy~\citep{Kaur2017EntropyMl}.

\subsection{Review Conclusions} \label{review_conclusions}
\subsubsection{Research Maturity} \label{research_maturity}
The sheer (low) quantity of works produced in the topic, relative to the many applications to several other areas of science that we observed during the review process, is itself an indicator of relatively \textbf{low maturity of research} in source code measurement through information theory. Of the 16 publications that attempted to perform entropy-based measurements of source code, five were purely theoretical, 10 were empirical, and one was a survey of the status quo in the 1990s. 

\subsubsection{Extent of Empirical Validation} \label{extent_empirical_validation}
Out of the 10 empirical works we assessed, only two attempted to validate their hypotheses via thorough empirical assessments. The other 8 used toy projects or code produced by students, often relying on external/programmer input, such as debug time, perceived complexity, or bug logs to assess their measurements. Although researchers since Hassan~\citep{hassan2009predicting} have started investigating the evolution of entropy throughout a project's life cycle, they rely exclusively on metadata, thus not considering the actual information contained in the source code files and their multiple levels of representation. None of the studies validated their metrics on samples of more than 6 projects, only one study attempted to measure the evolution of entropy metrics of graph-based representations of source code for small programs~\citep{Akundi2018CFG}, and 5 of the 10 papers assessed relied on code produced by students.

\subsubsection{Research Opportunities} \label{research_opportunities}
The relative scarcity of works that measured source code entropy and its evolution, coupled with the fact that most of the papers published in the field provided restricted empirical validation, leads us to believe that many of the conclusions at which Khoshgoftaar et al. reached when surveying the topic in the 1990s~\citep{Khoshgoftaar1994ApplicationsOI} still hold true. In particular, some foundational questions do not yet have satisfactory answers. Some of them are:
\begin{itemize}
    \item Do information theoretic measurements of source code capture the same dimension of complexity as conventional software quality/complexity measures?
    \item Can we obtain data from industrial size software systems to support the intuitive notion that a higher information content yields a higher quantitative/qualitative complexity?
    \item Can entropy-based approaches to measure source code complexity assist us in the process of source code management?
\end{itemize}

We believe that with the current compute power at our disposal and the availability of both open source code bases as well as tooling to process it, it is possible to conduct historical analyses of source code entropy metrics, identify their patterns of evolution, and arrive at satisfactory answers to our research questions, as well as to lay the groundwork for the development of tools that assist the management of complexity as a code base evolves.

\section{Methods} \label{study_settings}
In this section we describe the criteria for selecting projects, the tooling used to collect the data, and the kind of historical data (and metadata) that we collected for each change event. All of the source code for data collection, parsing, processing and analysis are available in the supplementary material \citep{supp_material}, as well as manual point analysis, visualizations and charts and inter-rater agreement measurements. For further explanation of how the material is organised, please refer to its \texttt{README.me} file in the root folder of the data repository.

\subsection{Data Collection}
To measure how the entropy of the source code evolves, we considered every commit in the history of the project. In practice, this means taking snapshots of the state of all source code files before and after each change is applied. We believe that the performance penalty that is incurred by traversing the history on a commit-by-commit basis, as opposed to applying arbitrary slices~\citep{canfora2014changes, keenan2022investigation}, is justified, as we are able to build an accurate picture from the shortest time interval possible: the time between any two consecutive commits. Our study is, to the best of our knowledge, the first one to track the entire commit history of projects whilst collecting their structural and textual entropy metrics, as well as the first to detect unusual change events via information theory.

We used the GitHub Search API~\citep{Dabic:msr2021githubSearch} to select our projects. Considering that our goal is to select projects that are actively maintained and have a relatively long history of change events our search query is defined by the following parameters:
\begin{itemize}
    \item Over 1000 (93th percentile of the 106213 Java projects returned by the GitHub Search API) commits merged to the project's main branch. Due to the running time of our serial process in a laptop with a 4-core, 8-thread Intel\textsuperscript{\textregistered} Core\textsuperscript{\texttrademark} i7-7700HQ at 2.8GHz with 32GB of memory, running the analysis for over 50000 commits took over five days, so we capped at that number;
    \item At least one commit in the last month;
    \item Project is not a fork;
    \item Project has over 1000 stargazers;
\end{itemize}

We restrict our discussion to a single programming language - Java - to simplify the engineering process of processing and parsing source code. Since we are analysing code both as a stream of tokens as well as parse tree nodes, we could analyse code written in any high level language in a similar way. Given the inherent naturalness of software \citep{hindle2016naturalness} and the fact that our information-theoretic definition of token and AST entropy is language agnostic, we believe our results are not impacted by choice of language. Given that the Shannon entropy needs only the relative frequencies of the symbols involved in the communication process, and that the Abstract Syntax Tree (AST) nodes producing during scanning and parsing are assigned types that correspond to the production rules in the grammar of the respective language \citep{aho86dragon, engineering_a_compiler}, we can use the frequencies of these node types to compute the structural entropy of programs written in any level language that is parsed into an AST. We could, for example, use the ANTLR parser generator \citep{parr2013definitive} to build parse trees for source code written in multiple languages and measure their Shannon entropies. As of the time of this writing, ANTLR is able to parse 168 languages \citep{antlrGrammars} and parses 47 out of the 50 most popular programming languages in 2024, including the 27 most popular \citep{solanguages2024}.

When computing token-based entropy, we calculated the total entropy both with and without code comments. Comments are an integral part of the source code, offering context, explanations, and insights into developer intent that the code alone does not capture. By estimating the entropy of comments in isolation and comparing them with the total token entropy, we can get a sense of the degree of naturalness of the code base. The rationale behind our decision to measure that came from manual inspection of projects where the code was mostly low-level or abstract and some of the most important files and classes contained mostly natural language comments and only a few lines of actual code. Measuring the effect of comments in isolation and the effect of their removal can also provide us with a metric of how informative or redundant the comments are. All of the projects we analysed contained comments to varying degrees, which we did not quantify during our experiments.

This query effectively filtered the total number of about 90000 projects written primarily in Java to a list of 562 candidate repositories. We drew projects randomly, excluding entries that did not correspond to actual production programs (books, courses and tutorials) until we arrived at 95 repositories that were actual software systems. The entire list of projects is available in the supplementary material \cite{supp_material}

\subsection{Tooling}
The application we built to process historical data from the projects has two main modules. The core module is written in Python and builds on top of the PyDriller framework~\citep{PyDriller}, which abstracts away Git functionality, thus allowing us to traverse the projects' histories chronologically via simple function calls. The second module, written in Java, is invoked by the first one, and uses JavaParser~\citep{javaparser} to generate histograms of AST nodes and edges from the programs' source code. When considering source code as a stream of words, we opted for ways to divide the code in as close a manner as possible to how words appear in natural language. For example, we wanted a function named \texttt{createUser} to be broken down into two separate words: \texttt{create} and \texttt{user}. For this we used the Spiral library~\citep{Hucka2018spiral}, which specialises in splitting such identifiers in source code files.

\subsection{Process} \label{methodology}
While performing a historical traversal of the entire commit log of the projects, we generate two data sets:
\begin{itemize}
    \item Commit level: contains commit metadata regarding the date, author information, message, total file count, message count, line insertions and deletions;
    \item File level: contains file path, type of modification (new file, update, rename, deletion), number of lines introduced or removed, total line count, token count, cyclomatic complexity~\citep{mccabe1976complexity}, and number of methods changed. We also apply equations \ref{h_token} and \ref{h_astedge} to compute the entropy metrics of each file. As the commits are traversed, not only do we compute the entropy of a given file at that point in time, but we also compare it to its previous version, so as to know how the entropy of each file is evolving over time.
\end{itemize}

To assess how our data assists us in answering \textbf{RQ 1}, we produce, for each project, a dataset of all changes to all files in each commit applied to the project's main branch. Upon aggregating the data for each commit, we generate a discrete time series where the horizontal axis corresponds to the timestamp of each commit analysed, whereas the vertical axis contains the total entropy of the project's source code at that point in time. In this case, the variation of the changes between two consecutive commits results from the changes to all individual files that were modified in the most recent commit. For structural entropy metrics, we collect AST edge entropy. For textual entropy, we apply equation \ref{h_token} to the entropy of all tokens (delimiters and operators discarded) and we also measure the effect of removing keywords and numbers. Natural language comments are considered both in isolation (i.e., disregarding the source code itself), as well as in conjunction with non-comment tokens.

Our search for answers to \textbf{RQ 2} consists in building correlation tables between classic code complexity metrics and our entropy-based ones, both at a file level as well as a commit level. The classic metrics collected were insertions, deletions, number of lines of code (nloc), changed methods count and insertions minus deletions, as well as McCabe's cyclomatic complexity~\citep{mccabe1976complexity}. We then correlate such metrics with structural and textual source code entropy.

\textbf{RQ 3} arises from a data-centric assessment of postulates regarding the evolution of software systems. Lehman put forward five laws which he claimed to govern software evolution~\citep{Lehman1980Laws}. Whilst admitting that software engineering is a human activity subject to a high degree of mutability and that one should not expect its evolution laws to be as precise as laws of physics, Lehman conducted a limited evaluation of these laws against real-world programs. Although some of the laws were verified against a limited set of programs, the context in which the laws were postulated is different from the current software engineering practices in a few ways. For example, the eight programs considered in their study were proprietary, which led to conclusions which may or may not hold in an open source development environment. Moreover, no more than eight programs were considered when the laws were postulated, and some laws were assessed on a sample of only three projects. In this work, we assess how our information-theoretic perspective of software evolution, upon mining 95 repositories, compares with Lehman's findings, thus answering \textbf{RQ 3}.

Our last question is \textbf{RQ 4}, which concerns itself with the applicability of changes in projects' information content as a way to detect unusual events. To this end, we perform anomaly detection on the time series generated in RQ 1 with respect to the total variation in entropy as a result of each commit, and investigate a parameter that can affect the detection of such events. With this in place, we then move on to analysing a statistically significant sample of such events, selected randomly, and inspect them manually to determine categories to which these events belong. We calculate false-positive rates to investigate whether such detection can be useful in practice.

\section{Empirical Assessment} \label{empirical-assessment}
The code that processes the historical data, the scripts that were employed to assist our inquiry, as well as the entire set of data frames, tables, and graphs that guided our assessment can be found at \cite{supp_material}. All the data was collected on a laptop with a 4-core, 8-thread Intel\textsuperscript{\textregistered} Core\textsuperscript{\texttrademark} i7-7700HQ at 2.8GHz with 32GB of memory. Further engineering efforts towards scalability and parallel execution, which could enable the analysis of larger samples of programs, is subject of future work. Section \ref{evolution_patterns} address \textbf{RQ 1}, whereas section \ref{correlations-classic} covers \textbf{RQ 2}. Section \ref{laws_of_evolution} and its sub-sections contain the results of our data-orientated assessment of each of Lehman's laws of software evolution, thus providing answers to \textbf{RQ 3}. We conclude our assessment in \autoref{outlier_detection}, where we lay out and evaluate answers to \textbf{RQ 4}.

We analysed a total of $1,827,204$ change events across $294,028$ commits.

\begin{table}
    \centering
    \begin{tabular}{lrrrrrr}
        \hline
        & \textbf{Contributors} & \textbf{Skew} & \textbf{Commits} & \textbf{Merges} & \textbf{Ratio} & \textbf{Files Changed} \\
        \hline
        Mean   & 222.03 & 8.66 & 4442.29 & 569.79 & 0.15 & 19233.73 \\
        Std    & 214.45 & 3.69 & 4828.56 & 819.13 & 0.14 & 24102.26 \\
        Min    & 2.00 & 0.00 & 8.00 & 0.00 & 0.00 & 1287.00 \\
        25\%   & 93.50 & 5.82 & 1649.00 & 82.50 & 0.04 & 5070.50 \\
        50\%   & 146.00 & 7.98 & 3076.00 & 223.00 & 0.09 & 9736.00 \\
        75\%   & 254.00 & 11.38 & 5175.50 & 746.00 & 0.22 & 19826.00 \\
        Max    & 941.00 & 18.58 & 29153.00 & 4235.00 & 0.53 & 139430.00 \\
        \hline
    \end{tabular}
    \caption{Summary of commit information}
    \label{tab:commit-stats}
\end{table}

\subsection{Patterns of entropy evolution (RQ 1)} \label{evolution_patterns}
\begin{figure}
    \centering
    \captionsetup{justification=centering}
    \includegraphics[width=\textwidth]{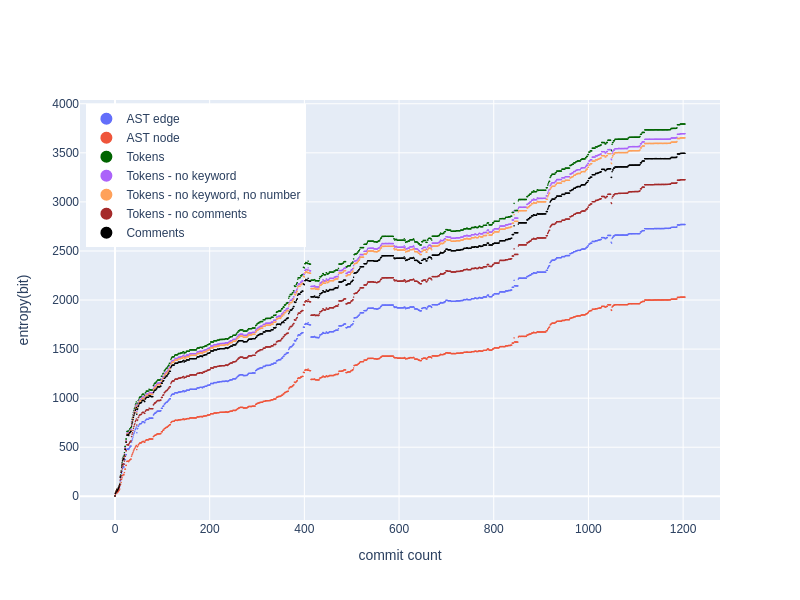}
    \caption{Typical time series of entropy evolution. Commits are presented according to their ordering in version control.}
    \label{fig:entropies-jib}
\end{figure}

Figure \ref{fig:entropies-jib} displays typical curves of evolution of structural and textual entropy. All but two projects (98\%) in the sample exhibit increasing trends as more commits are pushed to the mainline branch. As we shall see in further detail in \autoref{correlations-classic}, textual and structural entropy are usually strongly correlated, which is why all the curves in the figure appear to be vertical offsets of each other. Nevertheless, point-by-point inspection of textual and structural entropy distributions shows that their respective gradients are frequently in opposite directions (i.e., for several commits, an increase in structural entropy was accompanied by a decrease in textual entropy, or vice versa). It is worth noting that, for all projects, structural entropy metric scores are lower than the textual ones. This can be explained by the fact that our structural entropies are calculated based on Abstract Syntax Tree node types, which causes a lot of the textual information that is present in the source code to not be considered during structural entropy calculation.

\begin{figure}
    \centering
    \captionsetup{justification=centering}
    \includegraphics[width=\textwidth]{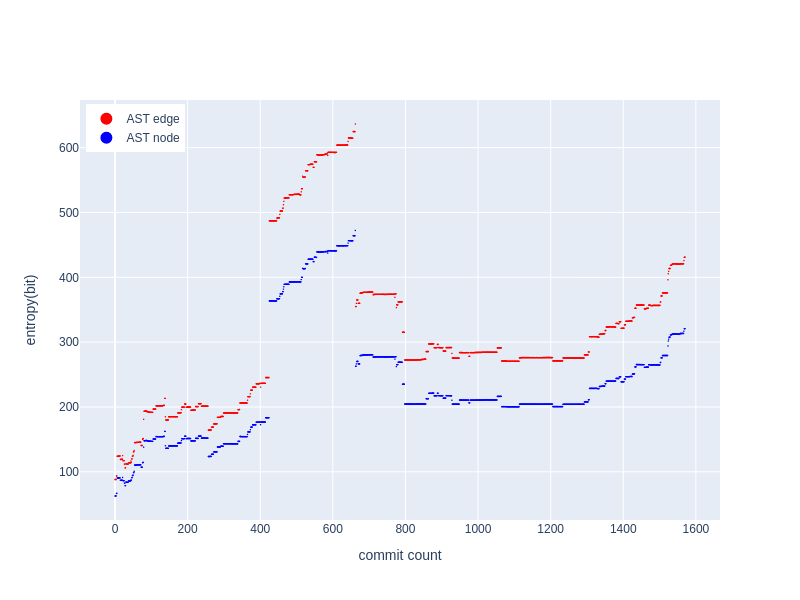}
    \caption{Evolution plot with big spikes and drops. Textual entropies are omitted for the sake of clarity. Commits are presented according to their ordering in version control.}
    \label{fig:entropies-ast-hikaricp}
\end{figure}

Although almost all projects share this trend, the smoothness of evolution can change drastically between projects. Figure \ref{fig:entropies-ast-hikaricp}, for example, shows the cases where entropy increases of 49\%, with a decrease of 79\% later. Table \ref{spikes-drops} summarises the absolute frequencies of projects that display spikes and drops of several percentages.

\begin{table}
    \centering
    \captionsetup{justification=centering}
    \begin{tabular}{cccccc}
     \toprule
          & 10\% & 20\% & 50\% & 80\% & 100\%\\ \midrule
    spike & 85 & 68 & 34 & 20 & 10\\ \midrule
    drop  & 75 & 53 & 34 & 24 & N/A\\ \bottomrule
     
    \end{tabular}
    \caption{Number of projects, (N = 95) which displayed increasing percentual changes of information content.}
    \label{spikes-drops}
\end{table}

Separation of concerns is a paramount concept in software engineering at various levels \citep{tarr1999nSeparation} and throughout all stages of its evolution \citep{ossher2001usingSeparation}, with effects ranging from cognitive complexity to software security \citep{de2002importanceSeparationSecurity}. Since entropy is based on the frequency of the elements of language analysed, a high density of information per file can indicate that the files contain too many different concepts, which can indicate that the project is handling multiple concerns. To assess how different projects handle this aspect of software evolution, we divided the total entropy at any point in time by the amount of source code files over which the entropy was distributed, so as to get a measure of entropy spread per file.

\begin{figure}
    \centering
    \captionsetup{justification=centering}
    \includegraphics[width=\textwidth]{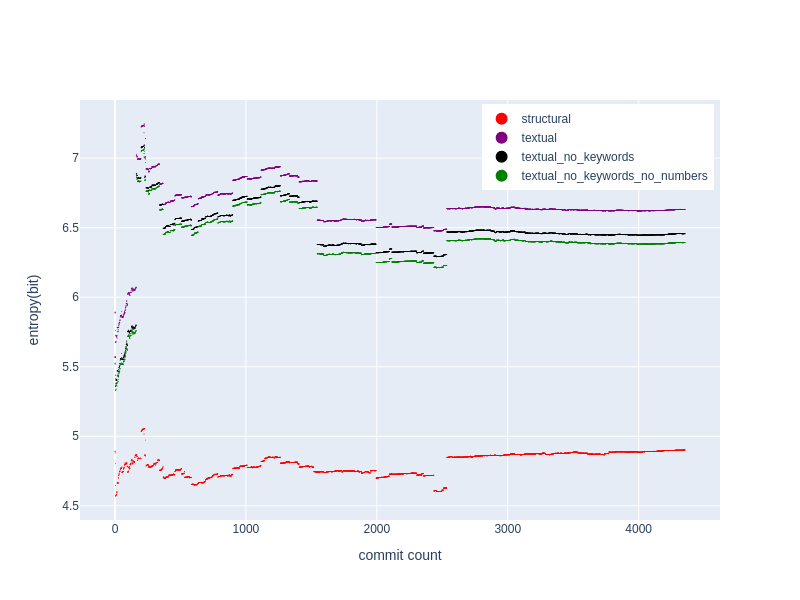}
    \caption{Project that keeps information spread stable throughout its lifetime. A single project presented this trend. Commits are presented according to their ordering in version control.}
    \label{fig:grpc-entropy-spread}
\end{figure}

\begin{figure}
    \centering
    \captionsetup{justification=centering}
    \includegraphics[width=\textwidth]{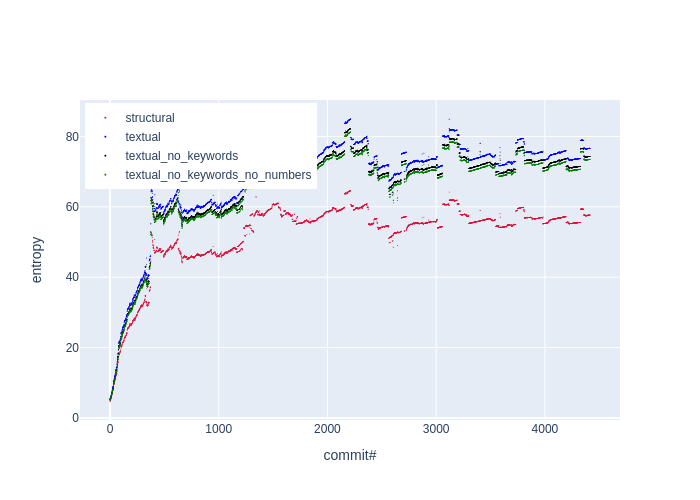}
    \caption{Initially increasing density of information per file, followed by stabilisation (76\% of the projects).}
    \label{fig:redisson-entropy-spread}
\end{figure}

\begin{figure}
    \centering
    \captionsetup{justification=centering}
    \includegraphics[width=\textwidth]{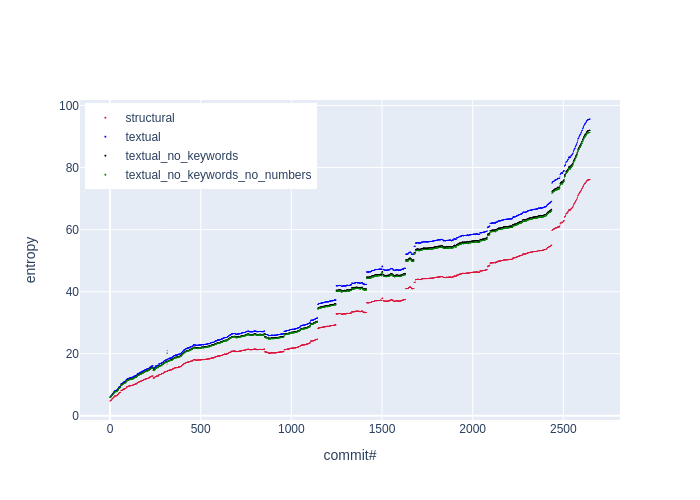}
    \caption{Ever-increasing information density per file (22\% of projects).}
    \label{fig:fresco-entropy-spread}
\end{figure}

We observed three main patterns:

\begin{itemize}
    \item Entropy spread remains at a low, almost constant value throughout almost the entire life cycle: This indicates that, even though the project's total information content is ever increasing, as more concepts are added to it, new files are introduced at equivalent pace so that the concerns are distributed more evenly over multiple files. Figure \ref{fig:grpc-entropy-spread} depicts this case. A single project displays this trend;
    \item Total information per file increases during early stages of the project, followed by stabilisation: This may suggest that, in its beginning, the main focus of the maintainers was to produce the desired functionality to be performed by the application, and, once the pace of evolution slowed down, more focus was placed on spreading the entropy throughout the code base. Figure \ref{fig:redisson-entropy-spread} depicts this example. This was the most common case, comprising approximately 76\% of the projects (n = 73) of the sample of projects;
    \item Entropy spread keeps increasing, similarly to that of the total information content of the projects. Figure \ref{fig:fresco-entropy-spread} exemplifies this case, which was common to about 22\% of the projects (n = 21). This suggests a development process where the growth amount of information introduced exceeds that of the creation of new files, which may mean files are handling multiple concerns.
\end{itemize}

\begin{tcolorbox}[colback=blue!5!white,colframe=black!75!black,title=RQ 1: What are the trends of evolution of source code entropy of actively maintained projects?]
\begin{itemize}
    \item \textbf{RQ 1.1}: What are the different patterns in the evolution of source code entropy?
    \item \textbf{RQ 1.2}: What are the similarities and differences in the trends of evolution of distinct projects?
\end{itemize}
The total amount of information has an upward trend for all projects. Even though two projects display a decreasing trend over its most recent 20\% commits, applying linear regression to all project's total entropy creates lines with positive slopes. While this was a consistent pattern for overall information content, the management of information on a per file basis reveals three main different patterns of information spread, from strict (one project) to initially uncontrolled, but eventually stabilising (73 projects), to ever increasing (21 projects).
\end{tcolorbox}

\subsection{Relationship between entropy and classic code complexity metrics (RQ 2)} \label{correlations-classic}
We computed the Spearman correlation between our two structural entropy metrics and four textual entropy ones---with varying degrees of tokenisation---with the metrics listed in \autoref{methodology}. The analysis was conducted both at a file level, as well as the commit level. This section describes the trends observed in the correlation tables generated for these two levels. 

Table \ref{tab:file-correls} contains a summary of the correlations calculated between entropy and each metric of code complexity. Correlations with nloc, cyclomatic complexity, token count and changed methods are weak. Insertions also display a moderate positive correlation on average, whereas deletions usually display week negative correlation, but with the widest range; for one project, there is a strong correlation of $-0.78$, and 38 projects have a moderate correlation (between $-0.40$ and $-0.70$). It is worth noting that the maximum absolute values of the correlations between insertions and deletions happens for the same project. The strongest correlation is that between entropy and $(insertions - deletions)$, which is strong ($\geq 0.70$) for three projects, and moderate for 71.

\begin{table}
\centering
\captionsetup{justification=centering}
\begin{tabular}{cccccccc}
\toprule
  & insert & delete & insert - delete & nloc & McCabe & tokens & changed methods \\ \midrule
  min & 0.25 & -0.78 & 0.33 & -0.01 & -0.05 & -0.32 & -0.04 \\ \midrule
  max & 0.81 & -0.10 & 0.87 & 0.38 & 0.36 & 0.03 & 0.37  \\ \midrule
  avg & 0.41 & -0.38 & 0.61 & 0.14 & 0.13 & -0.16 & 0.10 \\ \bottomrule
\end{tabular}
\caption{Spearman correlation (95\% confidence interval, p = 0.005) between entropy and classic metrics at the file level.}
\label{tab:file-correls}
\end{table}

In table \ref{tab:commit-correls}, the results are computed at the commit level. Note that the averages and ranges remain mostly preserved, with all but one value drastically changing: at this level, the correlations with deletions become weaker, with no projects exhibiting more than weak correlations.  
\\
\begin{table}
\centering
\captionsetup{justification=centering}
\begin{tabular}{ccccccccc}
\toprule
  & insert & delete & insert - delete & nloc & McCabe & tokens & changed methods \\ \midrule
  min & 0.22 & -0.25 & 0.26 & -0.03 & -0.05 & -0.03 & 0.10 \\ \midrule
  max & 0.77 & 0.30 & 0.87 & 0.36 & 0.32 & 0.34 & 0.56  \\ \midrule
  avg & 0.41 & 0.00 & 0.51 & 0.11 & 0.10 & 0.11 & 0.27 \\ \bottomrule
\end{tabular}
\caption{Correlations between entropy and classic metrics at the commit level}
\label{tab:commit-correls}
\end{table}

\begin{tcolorbox}[colback=blue!5!white,colframe=black!75!black,title=RQ 2: How do information theory-based metrics of source code entropy correlate to standard metrics of software complexity?]

The correlation between structural and textual entropy varies: 2 projects displayed weak correlations, 61 projects displayed moderate, and 32 had strong correlation. There is a \textit{low}, and \textit{at best moderate} correlation between our entropy measurements and, insertions, deletions, nloc, tokens and changed methods. The correlation with McCabe's cyclomatic complexity is remarkably low, ranging between $-0.05$ and $0.32$.
The correlation between entropy and $insertions - deletions$ was the strongest on average, but this was also highly variable within the projects.

The above findings suggest that it may be possible to measure different aspects of code complexity via the entropy of representations of source code.
\end{tcolorbox}

\subsection{Entropy and laws of software evolution (RQ 3)} \label{laws_of_evolution}
In this section, we investigate how information-theoretic measurements of software evolution via entropy relate to laws of software evolution formulated by Lehman during the early 1980s~\citep{Lehman1980Laws}. Whilst it is important to acknowledge that software engineering processes are subject to human judgement and decision-making, and that therefore their degree of precision cannot be the same as laws of physics~\citep{Lehman1979Understanding}, we believe that a quantitative approach to the evaluation of such laws can both reinforce the validity of the aforementioned laws as well as help us reassess some of them when suggested by data. Lehman's original work based itself on no more than eight---sometimes only three---programs to arrive at his laws, which are widely accepted in the field. No explicit details about what these projects are can be found. Moreover, these laws were proposed at a time where the landscape of software development was drastically different from today's, ranging from the availability of compute power and the speed of spread of software engineering knowledge to the wide availability of open source projects that are maintained by voluntary contributors, some of which being nonetheless essential to other open source projects, as well as to proprietary software of big enterprises. We evaluate each law based on the data of the 95 projects surveyed in this work. Each of the following subsections map to Lehman's laws in order of appearance in his publication~\citep{Lehman1980Laws}.

\subsubsection{Continuing change} \label{continuing_change}
The first law states that the inherent malleability of the programming process is such that, in order to remain useful, applications must undergo continuing change. As discussed in \autoref{evolution_patterns}, none of the projects considered in our work display constant total information content. Although projects may contain periods where overall entropy remains relatively stable, all systems are subject to constant changes, which is reflected by the variation of the entropy metrics. Virtually all commits that modify source code files introduce or remove information from the projects, which leads us to conclude that our entropy-based measurements are generally in \textbf{agreement} with this law.

\subsubsection{Increasing complexity} \label{increasing_complexity}
\begin{figure}
    \centering
    \captionsetup{justification=centering}
    \includegraphics[width=\textwidth]{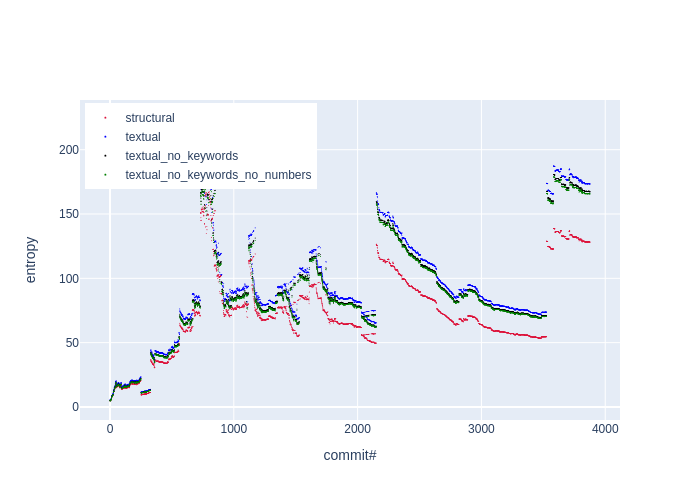}
    \caption{An increasing entropy spread over time, albeit certain long ranges commits attempted to reduce it.}
    \label{fig:skywalking-entropy-spread}
\end{figure}

As its name suggests, this laws states that a natural byproduct of the first law is that the complexity of programs, which serves as a proxy for its deteriorating structure, is bound to increase unless engineering efforts are directed towards controlling it. Our data corroborates with this law in three ways:
\begin{itemize}
    \item The net information content of $98\%$  ($n = 94$) of the projects evolves in such a away that the total amount of information displays increasing overall trends. Although one can select arbitrary time windows during which the total entropy decreases, the analysis of the entire history of the projects strongly agrees with the law of increasing complexity;
    \item As can be seen in table \ref{spikes-drops}, it is not uncommon for projects to display, between two consecutive commits, sudden increases in the total entropy of the code base;
    \item We saw in \autoref{evolution_patterns} that about 21\% of the projects not only show an increase in net entropy, but also showed an increase in amount of information per file. Although all but one of the remaining projects eventually controlled the information spread, this would usually not start happening until 30 to 50\% of the development of the project has developed. Moreover, graphs like the one in Figure \ref{fig:skywalking-entropy-spread} are common throughout; there are usually periods of increasing density of information across files, followed by commits that attempt to distribute information better. Still, the trends are toward increasing amount of information per file.
\end{itemize}

Our analysis of the information-theoretic data produced from our experiments suggests that its measurements are mostly in \textbf{agreement} with the Law of Increasing Complexity. Furthermore, pictures like Figure \ref{fig:skywalking-entropy-spread} reinforce what the law states, that is, directed engineering efforts are necessary to keep complexity under control.

\subsubsection{Fundamental law} \label{fundamental_law}
Lehman's third postulate is the fundamental law of program evolution, or the \textit{Law of Statistically Smooth Growth~\citep{Lehman1979Understanding}}, and states that software engineering dynamics are such that the programming process itself, as well as global and system attributes measures are self-regulating, and that they display determinable trends and invariants which ensure the project evolves in a stable fashion. We found the following aspects to be universal to most projects.

Lehman recognises that this law may be subject to the state of the art of software development of the respective era in which software is developed, but that the organisational dynamics of the projects they analysed ensure statistically smooth growth~\citep{Lehman1980Laws}. Whilst almost all projects observed display increasing growth over time, results such as those of table \ref{spikes-drops} and by our measurements of entropy evolution---like the ones in figures \ref{fig:entropies-jib} and \ref{fig:skywalking-entropy-spread}---indicate that, from an information-theoretic perspective, the growth of the open source projects herein considered tends to not be smooth or stable; our data shows that it is not uncommon for two ranges of commit to display drastically different patterns (i.e., 200 commits where entropy remains mostly stable, as well as 50-commit ranges where entropy doubles and then drops by $10$ to $50\%$).

Moreover, Lehman states that "there \textit{exists} a dynamics whose characteristics are largely determined during the conception and early life of the system of the maintenance process and of the maintenance organization. The characteristics of this dynamics increasingly determines the gross trends of maintenance and enhancement projects"~\citep{Lehman1979Understanding}. Our data suggest that virtually all projects have two separate phases: one in which the entropy increases at a much higher pace, during its early stages, followed by a phase where even though entropy is still increasing, it does so at a slower pace. Moreover, an analogous pattern is present for information spread per file. This goes against the idea that early-phase decisions strongly shape the process.
Since this is in stark contrast to the concept of smooth evolution, our data from open source projects is mostly in \textbf{disagreement} with this law. It is important to acknowledge, though, that not only did Lehman predict that this would depend on the state of software engineering practices at the time of analysis, but his work was mainly done on privately held and managed projects. Our context of open source projects, which presents a challenge to coordination and efficiency, may be the driving factor in such disagreement.

\subsubsection{Conservation of organisational stability} \label{conservation_stability}
This law, also called by Lehman the invariant work rate, postulates that organisations seek stable growth, as well as internal stability. As a result, Lehman writes that "sudden substantial changes in managerial parameters such as staffing, budget allocations, manufacturing levels and product types" are to be avoided, if not made impossible altogether. The author also states that organisational checks and balances ensure smooth progress as projects evolves The way this was measured was by counting the number of modules changed per unit of time, and although no further measurements were provided, Lehman asserts that this measure remained statistically invariant over time.

Although Lehman's writings do not make explicit what kind of projects/organisations were surveyed by their work, it appears that the scope of their research was that of private corporations, wherein one would expect such hypotheses to hold true. However, the current landscape of software development is not only influenced by open source projects, but also involves startup companies, wherein growth may be both fast and unstable, and is often dependant on extraneous factors, such as competition, availability of resources, and macroeconomic factors~\citep{graham2012startup}. Moreover, also in similar fashion as the previous law, many software projects available today---and all of the 95 projects we consider in this research---are open source, which, by their nature, do not strictly enforce organisational checks and balances as the ones present in the private sector. This may result in different patterns of evolution than those of privately developed systems.

From an entropy perspective, we argue in a similar fashion to \autoref{fundamental_law}; our results tend to disagree with Lehman's fourth law. As seen in \autoref{evolution_patterns} and in table \ref{spikes-drops}, coupled with the voluntary nature of contribution to open source projects, the projects assessed in this work seem, in general, to not display such organisational stability, so our data is in relative \textbf{disagreement} with the law.

\subsubsection{Conservation of familiarity} \label{conservation_familiarity}
The fifth and last of Lehman's laws, to which he also referred to as the Law of Incremental Growth limits, states that "the incremental growth of the program varied widely from one release to the next, but the average over a relatively high number of releases remained remarkably constant". This implies that, usually, releases that present very high growth tend to be followed by releases containing little or no growth. However, no data, measurements or methods on how the three systems that were considered when arriving at this law were explicitly provided. The author himself acknowledge that, at that time, there was no precise way of measuring the total content of a releases. Due to the lack of specificity in the statement of the law, its explanation and interpretation become difficult. Also, our approach was on a commit-by-commit basis and that we did not seek ways to automatically determine when major releases happened for all of the 95 projects herein considered, we decided to defer the discussion of this law to future work.

\begin{tcolorbox}[colback=blue!5!white,colframe=black!75!black,title=RQ 3: To what extent do information-theoretic measurements of source code evolution via entropy agree with Lehman's laws of program evolution?]

Our entropy-based measurements of source code evolution agree with two laws: continuing change and increasing complexity. Such laws are general aspects of software evolution and are less sensitive to managerial settings. However, our data disagree with the two laws that are more sensitive to management approaches, namely the fundamental law and the law of organisational stability. Our data as well as our current open source development context does not support the hypotheses that there exist self-regulatory processes that ensure smoothness of evolution. Our data neither support nor contradict Lehman's fifth law.
\end{tcolorbox}

\subsection{Entropy-based Detection of Unusual Events (RQ4)} \label{outlier_detection}
The sources of information with which maintainers interact when evaluating source code changes are several: pull request discussions, commit messages, issues, and naturally the source code itself. Therefore, in order to manage their attention and use relevant information to make decisions, developers can benefit from tools that detect unusual events~\citep{Treude2015DevActivity}. To that end, we developed a process to capture events that, from the perspective of source code entropy, are unusual. As they can significantly impact the total information conveyed by the project, such events may introduce long term maintenance issues (i.e. bugs, inflexibility to change, unnecessary complexity), and must be carefully assessed by maintainers.

\subsubsection{Methods} \label{detection_methods}
To detect such events, we performed real-time detection of variations in the total entropy of the project files between consecutive commits. By traversing all commits in chronological order and keeping track of the deltas between the entropy before and after a commit is pushed, we are able to perform anomaly detection to identify changes whose total information inserted or removed from the project were outliers relative to the distribution of entropy changes until that point in time. We did this for all commits in sequence, therefore applying the smallest time slicing possible (i.e., maximum granularity). The parameter that was configurable in our detection was the maximum memory size, which defines how far back in the past, relative to the current commit, to keep track of; events that fall outside this window are removed from the distribution of deltas. For instance, if this value is set to 70\% over 1000 commits, then after commit $700$, the first events will start to be removed from the distribution of entropy deltas. This parameter effectively enables the detection to ``forget'' events that are too distant in the past. For our experiment, this parameter was configurable to be 50\%, 75\% or 100\% of the commits. For each possible configuration, we traversed the projects' commit histories and detected all events that were outliers with regard to information introduced or removed. We then randomly selected points for manual analysis. We considered a confidence level of 95\% to calculate the number of events that needed to be assessed ($n = 321$). Each event was analysed twice according to the following procedure:
\begin{itemize}
    \item Visit the commit page on the GitHub web user interface, read all textual information available (commit message, pull request page and discussions, related issues);
    \item For each of the aforementioned sources of information, research relevant terms that may be specific to the domain of the project;
    \item Consider number of lines introduced or removed, as well as the number of files;
    \item Categorise the event;
    \item Determine whether the event is a true positive or not, i.e., if the change would need to be prioritised by maintainers.
\end{itemize}

Events related to test, autogenerated, example, and outdated code were immediately labelled as false positives, as these do not impact the functioning of the resulting system. Since we are interested in significant introduction or removal of information, any remaining event whose variation of entropy was an outlier relative to its distribution. Since we are interested in unusual events and know the mean and standard deviation of the distribution, we used standard score (or z-score) to define unusual events. Any change that had a z-score less than or equal to 3 relative to the distribution of entropy deltas was also considered a false positive. Without proper knowledge of the domain as well conventions and expectations agreed upon by the projects' stakeholders, it can be difficult to qualitatively assess changes with lower z-scores. We considered all other events as potential true positives and categorised them according to the labels defined in \autoref{event-categories}. Any event that was a potential true positive after these criteria was manually inspected once again.

The first author conducted the initial labelling of statistically significant samples of events, and discussed partial results on weekly meetings with the other three authors. To validate the labelling of an unusual event as being important or not, the first and second authors drew a random sample of 50 unannotated events and applied the aforementioned procedure to decide whether the event was a true or false positive. We calculated inter-rate reliability using Krippendorff's alpha \citep{Zapf2016MeasuringIR} separately for each column, before and after a joint discussion to reconcile the assessments. Regarding the relevance of the event, the initial agreement was $\alpha = 0.61$, which improved to $\alpha = 0.87$ after discussion, whereas the respective coefficients for the categorisation of the events went from $0.78$ to $0.95$. Using Landis and Koch~\citep{landis1977measurement} strength of agreement levels, we observed the agreement regarding the relevance of the event go from substantial to almost perfect, while the agreement on categories remained almost perfect. We discuss the limitations of this process in \autoref{limitations}.

Due to the need to manually analyse a statistically significant sample and as an attempt to prevent the different kinds of bias when judging in uncertain conditions \citep{tversky1974judgment}, we opted for a conservative strategy when characterising the points. Not only did we discount points that did not fit our statistical criteria, but we also considered any points that matched our criteria but whose manual inspection did not provide substantial evidence for it being a true positive. We defined three progressive levels of strictness:
\begin{itemize}
    \item \textit{Maximally conservative}: Only points that correspond to insertions and updates of software and product features;
    \item \textit{Conservative}: Also includes events where software or product features were removed. Note that this does not correspond to removal of unused/deprecated features, but in fact to the decision to remove a feature, which does change the behaviour of the resulting system;
    \item \textit{Least conservative}: Includes change events whose changes are such that although the observable behaviour of the system may not have immediately changed, the change may still have long-term implications to the project. Events added fall into groups like removal of dependencies into separate repositories or software porting, either in the form of implicitly importing the entire code base of a dependency, as well as forks of stable versions of dependencies to allow for customisation of behaviour;
\end{itemize}

\subsubsection{Nature of Detected Events} \label{event-categories}
The procedures applied in the previous section led to the definition of the following categories:
\begin{itemize}
    \item \texttt{config}: changes to the project's build configuration files;
    \item \texttt{documentation}: commits that exclusively change the documentation of the system;
    \item \texttt{file\_ext\_convert}: conversion of files from one extension to another whilst preserving content;
    \item \texttt{ex\_code}: addition or removal of code\footnote{We remind the reader that, although we are referring to addition and removal of code from the project, technically we are concerned with the total information represented by the source code. For further details, please refer to \autoref{correlations-classic}.} that exemplifies how the project is expected to be used;
    \item \texttt{autogen\_code}: addition or removal of code that was introduced by code generators. We considered all change events that introduced automatically generated code as false positives regardless of the magnitude of the change, as developers have a lot less control over this kind of code;
    \item \texttt{soft\_feat}: introduction (\texttt{add}), improvement (\texttt{up}) or removal (\texttt{rem}) of features that seek to enhance the projects' software architecture, performance, or code that is consumed by other parts of the application (i.e., interfaces, indirection, abstraction layers, API, monitoring, performance improvements). This category also comprises porting events like forks of other repositories and extraction of dependencies into their own repositories. Can be followed by \texttt{add\_dep} or \texttt{rem\_dep} in cases where dependencies are introduced or removed from the build. \texttt{ext\_dep} is used when a dependency or module is extracted into its own software repository;
    \item \texttt{refactor}: changes to the internal organisation of the code base, whose goal is to facilitate understanding of the code. May or may not change behaviour \footnote{Note that this definition is not as strict as Fowler's definition of refactoring, which preserved behaviour \citep{fowler2018refactoring}. We decided to accept potential changes of behaviour because the term is often used by practitioners in a less rigorous way, which generally accepts refactors to alter behaviour.};
    \item \texttt{rebase}: Branch rebasing;
    \item \texttt{old\_code\_rem}: removal of deprecated and/or unused code;
    \item \texttt{test\_code}: insertion or deletion of test code;
    \item \texttt{bug\_fix}: correction of malfunctioning behaviour;
    \item \texttt{prod\_feat}: introduction, improvement or removal of features that are noticeable by end users (user interface changes, user-facing features, design);
    \item \texttt{commit\_revert}: reversal of a previously introduced commit;
    \item \texttt{multiple\_changes}: A combination of two or more of the above categories.
\end{itemize}

The sets of events described in Table \ref{tab:total-point-count} are not mutually exclusive. In fact, the events detected by the maximum memory configuration are almost a proper subset of the other two sets of events; only 4 data points are exclusive to the maximum memory setting, all four being true positives where software features were introduced (3) or updated (1).

\begin{figure}
    \centering
    \captionsetup{justification=centering}
    \includegraphics[width=\textwidth]{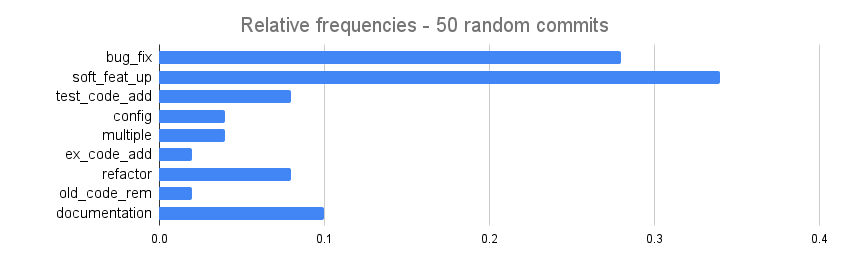}
    \caption{Categories for a sample of 50 random commits}
    \label{fig:50-random-points}
\end{figure}

\begin{figure}
    \centering
    \captionsetup{justification=centering}
    \includegraphics[width=\textwidth]{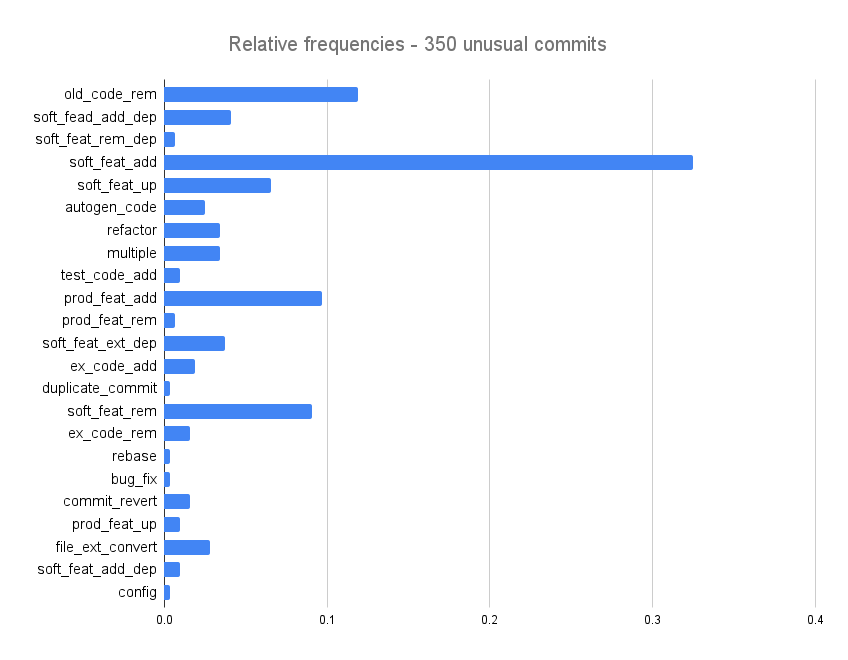}
    \caption{Sample of 320 points detected by full context window}
    \label{fig:full-context-categories}
\end{figure}

\begin{figure}
    \centering
    \captionsetup{justification=centering}
    \includegraphics[width=\textwidth]{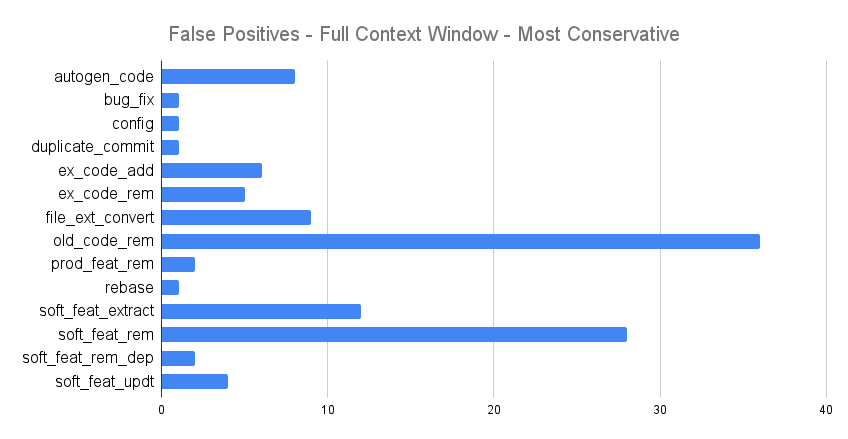}
    \caption{False positives for full context window}
    \label{fig:full-context-false-positives}
\end{figure}

Figure \ref{fig:full-context-categories} shows the distribution of categories for a sample of 350 points that were detected using a full context window. The most prevailing category is the addition of software features, with 104 events representing $32.5\%$ of the total. Removal of old, deprecated, or unusual codes is also common (n = 38). Categories related to removal of software features, as well as insertions, removals and updates of product features comprise most of the remaining distribution. Note how this distribution differs from the distribution in figure \ref{fig:50-random-points}, which contains 50 usual events. There is no addition of new features in this case; only updates to existing ones. In fact, none of the 50 points assessed in figure \ref{fig:50-random-points} had more than 200 lines changed. Note also how bug fixes and documentation updates are much more prevalent. This satisfies the intuition that day-to-day commits are simpler in nature. 

In Figure \ref{fig:full-context-false-positives} we depict the false positives of the full context window, which we obtained by manually inspecting each event listed in Figure \ref{fig:full-context-categories}. In our most conservative setting, removal of software features are not considered relevant events, so 30 of such removals feature as false positives. Apart from this category and four feature updates, all the remaining categories are events that in fact rank little in importance in terms of source code: configuration changes, conversion of file extension, dependency management, and addition or removal of example code comprise the vast majority of the false positives.

\begin{figure}
    \centering
    \captionsetup{justification=centering}
    \includegraphics[width=\textwidth]{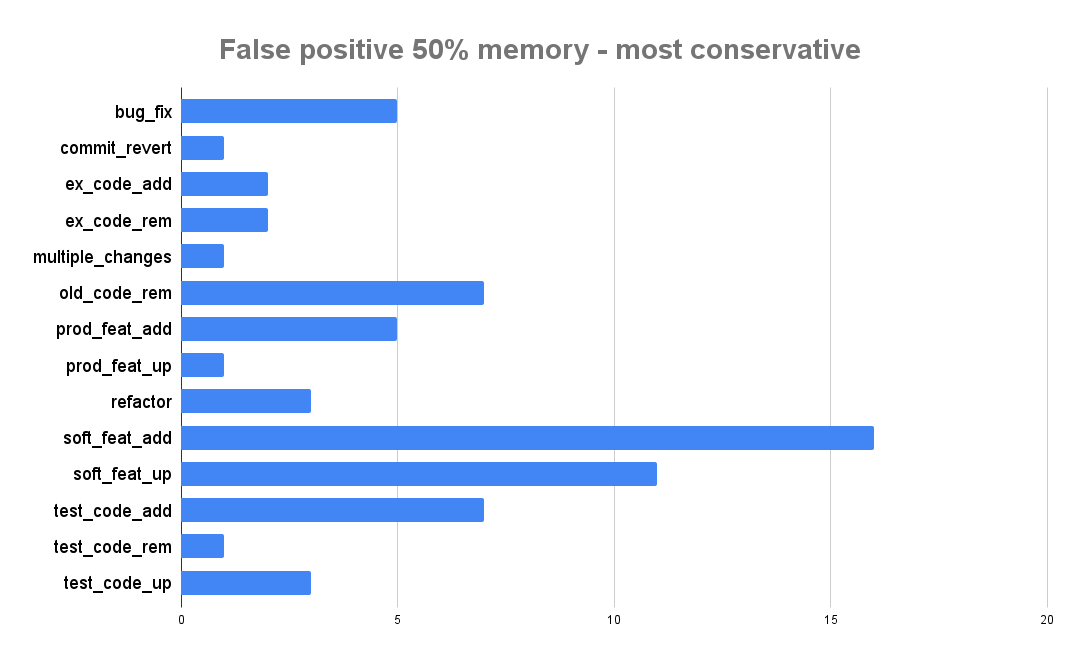}
    \caption{False positives exclusive to 50\% memory size}
    \label{fig:50-mem-false-positives}
\end{figure}

\begin{figure}
    \centering
    \captionsetup{justification=centering}
    \includegraphics[width=\textwidth]{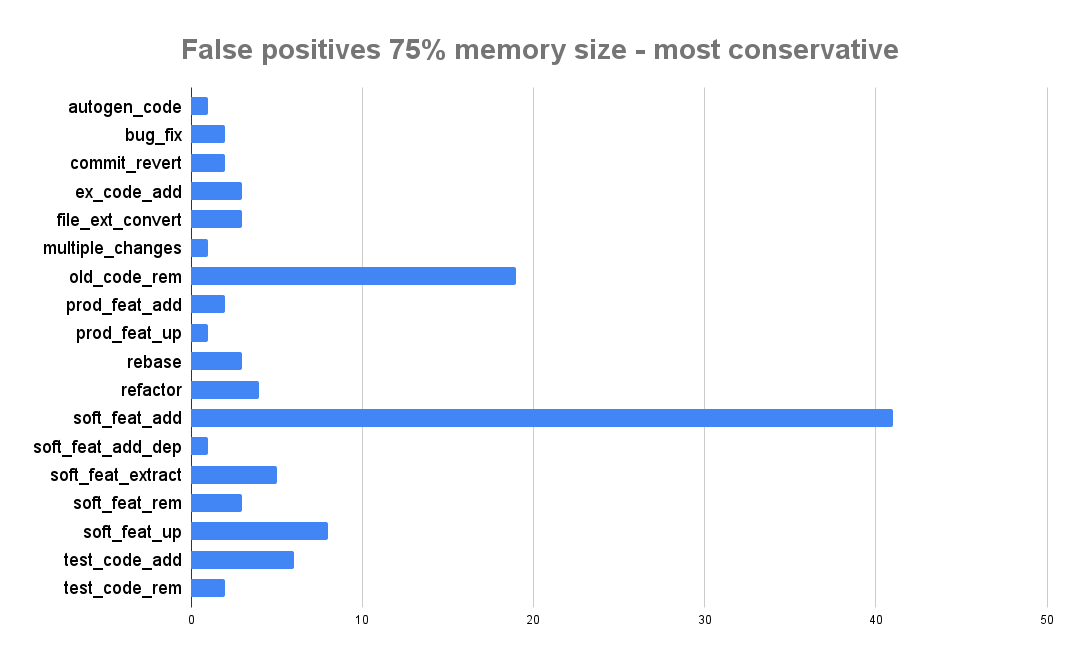}
    \caption{False positives exclusive to 75\% memory size}
    \label{fig:75-mem-false-positives}
\end{figure}

Figure \ref{fig:50-mem-false-positives} and Figure \ref{fig:75-mem-false-positives} show the distribution of false positives for the maximum memory of 50\% and 75\%, respectively. Notice how the distribution changes relative to Figure \ref{fig:full-context-false-positives}; reducing the memory size causes some software feature additions to be labelled as false positives to increase in both cases; 16 ($24.62\%$) and 41 ($32.94\%$). As described in \autoref{detection_methods} and depicted by Figure \ref{fig:entropies-jib}, later stages of a project's history tend to generate distributions of variations of entropy for whom the criteria defined in \autoref{detection_methods} of rejecting points whose entropy variation had a z-score less than 3, or changes whose actual feature code introduced corresponds to less than 50\% of the total entropy contribution, may turn out to be too strict. Integrating the information from the slopes of the segments encompassed by the window defined by a given maximum memory size into the definition of criteria for unusual events for a particular region of commits may cause many of the false positives detected for limited memory sizes to be properly labelled as important events, therefore improving the precision of the detection.

\subsubsection{Effect of Memory Size}
\begin{table}
\centering
\captionsetup{justification=centering}

\begin{tabular}{cccc}
\toprule
  & max memory 100\%& max memory 75\%& max memory 50\%\\ \midrule
  total points / 95\% C.I. & 1879 / 320 & 2225 / 328 & 2122 / 326  \\ \bottomrule
\end{tabular}
\caption{Total points and 95\% C.I. sample for a window that considers the 100, 75 or 50\% of the most recent commits}
\label{tab:total-point-count}
\end{table}

Table \ref{tab:total-point-count} summarises the total number of points whose entropy values were outliers, as well as the respective sample size for a 95\% confidence interval. A 75\% maximum memory detected the highest amount of points, whereas remembering all events produced the smaller set of events.

For each of the cells in Table \ref{tab:total-point-count}, we selected a random sample that matched the 95\% confidence interval value and applied the criteria outlined in \autoref{detection_methods} to classify the events. Table \ref{tab:percentages-varying-memory} displays the percentages of true positives for each level of strictness when we vary memory size. A total of 974 points were analysed. From these two tables, we can see that although a 75\% memory size captures the highest number of events, it is, at the same time, less precise than considering all events, but more precise than a shorter memory of 50\%.

For each configuration, we plotted the position of the detected points in the project's timeline, visited each point again, and observed the absolute value of the variation in entropy caused the respective events. For \textit{all} projects, we observed that the extra points captured by reducing the memory size occurred at later stages of the project and by points that were not detected when considering the full window. 

This is consistent with plots like Figure \ref{fig:entropies-jib}, where we can visually see that the slope of the evolution is much steeper in the beginning of the commit history, and that the process tends to slow down as the project stabilises\footnote{In fact, we split the time series generated for all projects and applied linear regression to each segment, and the steepest slopes were invariably in the beginning of the project.}. In other words, the concentration of outlier events that are true positives by our criteria is greater in the regions where the slope of the segments is higher, which are consistently the earlier segments. This means that reducing memory size causes the detection to ``forget'' the massive events that occur in the beginning of the project. As a result, these events are not considered when a limited memory size is applied and the project is in its later stages, thus yielding a higher rate of false positives, suggesting that the criteria of detection themselves may need to be different depending on the rate of change in entropy. As we shall see next, the distributions of false positive events also change.

\begin{table}
\centering
\captionsetup{justification=centering}
\begin{tabular}{cccc}
\hline
  & max memory 100\%& max memory 75\%& max memory 50\%\\ \hline
  most conservative     & 58.12  & 47.98 & 37.11  \\ \hline
  conservative          & 72.50  & 52.74 & 38.97  \\ \hline
  least conservative    & 83.13  & 59.14 & 40.12  \\ \hline
\end{tabular}
\caption{Percentages of true positives versus maximum memory size}
\label{tab:percentages-varying-memory}
\end{table}

\begin{tcolorbox}[colback=blue!5!white,colframe=black!75!black,title=RQ 4: What kind of change events can we detect using entropy-based anomaly detection?]
Almost \textbf{two thirds} of the events relate to features that are internal concerns of the system (\textit{programmer-facing complexity}). Typical events are performance/architectural improvements and tooling. Events that impact \textit{user-facing complexity} are relatively rare ($10.30\%$). Still, these events usually involve a large number of files being changed, and their potential impact on the system's usability can be particularly important to maintainers. Almost no events ($3.75\%$) are refactors or bug fixes. Among the false positives, the most prevalent event is removal of old/unused code ($31\%$).

Limiting the maximum window of events that comprise the distribution upon which we run the detection has the effect of increasing the number of detected events by about $20\%$. However, if the same criteria for false positives is maintained regardless of the memory size, we observe decreases in the precision of the detection.  The strictness of criteria for defining a false positive is sensitive to the recent history of the project and may be a defining factor as to whether automatic entropy-based detection of unusual events is practical.
\end{tcolorbox}

\subsection{Discussion and Implications}
Due to the nature of the curves of evolution, restricting the memory size can cause several important events to be labelled as false positives, if we apply the same level of strictness to events that fall in a high slope region versus a low slope region. Moreover, remembering all events may cause certain important events to not be captured at all during the later stages of the project, therefore hurting false positive rate.

Our conservative and least conservative true positive ratio are 72.50\% and 81.13\%, respectively. Given that practitioners tend to consider a precision of 75-80\% as a minimum threshold before adopting new tools, and that they "care much more about too many false positives than about too many false negatives" as well as that "high false positive rates lead to disuse"~\citep{Christakis2016DevelopersNeeds}, improvements to context sensitivity of the calculation of entropy, consideration of other statistical measures of surprisal (KL-divergence, mutual information, cross-entropy), and an awareness of the current pace of evolution of the project's information content may cause this way of detecting unusual events to yield acceptable precision and recall metrics.

We envision development of tooling that is able to maintain the historical changes to the total source code information of the project and warns stakeholders if unusual amounts of information were added or removed from the source code. Other potential areas of application are measuring the total information required to express different designs, to quantify the impact of refactoring, and to measure contextual cohesion.

This method of detection may be configurable to be optimal, i.e. initially all events are considered, and as the project evolves, we keep track of the pace of its evolution and use that to inform decisions regarding whether to start discarding earlier events, as well as to adjust the criteria for detecting unusual events.

We expect our work to be useful to the different parties that may be involved with software development activities in the following ways:

\begin{itemize}
    \item We hope to have made \textbf{researchers} aware of another approach to statically assess properties of source code, which may measure different dimensions of code complexity. We also aim to pave the way for quantitative measurements of the impact of source code changes and ways to organise code, such as refactoring efforts or when comparing two different ways to structure the solution of a domain problem;
    \item  We also encourage that \textbf{maintainers} be involved in this kind of empirical research, given that they are the stakeholders who probably deal the most with managing source code changes.
    \item  Finally, we invite \textbf{educators} and \textbf{developers} of all levels of experience to incorporate the view of programming as a communication system where the source code is one of the most used channels where stakeholders exchange information, and to consider the impacts that their changes may have in such channel.
\end{itemize}

\section{Implications: Automating Detection of Surprising Code Changes} \label{vision_automated_detection}

Automating the detection of surprising events in software development is motivated by the inherent complexity and continuous evolution of codebases, as well as by the multiple sources of information to which stakeholders are exposed. As software projects grow and involve multiple contributors, the frequency and volume of changes can make it challenging for maintainers to track all modifications effectively. Surprising events, such as significant deviations from established coding patterns or unexpected increases in code complexity, can introduce bugs, technical debt, or architectural issues that may go unnoticed during regular code reviews due to the amount of information maintainers need to process to make their decisions \cite{TREUDE2018237Unusual}. Using automation to detect these anomalies, developers can focus their attention on potentially problematic changes that require immediate intervention.

It is important to emphasise that the ideas presented here are intended as speculative opportunities for future research rather than as definitive claims. Our current study does not assert that entropy-based detection should replace existing static analysis tools; instead, it seeks to find ways to complement it by considering the evolution of the code base. Future studies could explore how integrating entropy metrics with traditional analysis methods could enhance the identification of subtle yet impactful code changes, thus enriching our understanding of software evolution. This research direction could help establish a theoretical basis for new automated tools that leverage information theory to complement conventional practices.

Shannon's information theory provides a robust mathematical framework to quantify the unpredictability or entropy within a system. In the context of software development, entropy can be used to measure the complexity and disorder introduced by changes in the codebase. Automated detection systems based on information theory can identify unusual entropy spikes, signalling that a change may have introduced a higher level of complexity or deviation from the norm. It is important to note that our approach does not aim to replace or outperform static analysis tools but rather to capture different aspects of software evolution. Specifically, static analysis focuses on identifying violations of coding rules and potential defects at a given point in time, whereas entropy-based detection highlights unusual evolutionary patterns by quantifying changes in information content over time. As demonstrated in our correlation analysis, surprising entropy variations do not necessarily align with static analysis warnings, reinforcing that these methods serve distinct but complementary purposes. These systems can then alert developers to these changes, allowing them to assess the impact and mitigate potential risks before the changes reach end users or escalate into more significant issues.

Moreover, automating the detection of surprising events helps maintainers manage the growing amount of information generated in a software repository. As the number of commits, pull requests, and code modifications increases, the cognitive load on developers also rises \citep{kalyuga2011cognitive}.

Information-theoretic metrics could be used in a way analogous to the Maintainability Index \cite{coleman1994metrics}, where Halmstead volume \cite{halstead1977elements}, cyclomatic complexity \citep{mccabe1976complexity} and lines of code (loc) are combined to produce a value that represents the complexity of the source code. They could even be integrated with classic metrics of complexity to result in a more encompassing metric.  We assume a synthetic metric that is mathematically well defined and draws is based on entropy - but can also integrate other definitions from information theory, such as cross-entropy \citep{hastie01statisticallearning}, Kullback-Leibler divergence \citep{KLdivergence} and mutual information \citep{kraskov2004estimating} - which we will call the \textbf{Surprisal Index (SI)}---whose normalised values range between 0 and 1---, which can be measured for three different scopes: function, file, and the entire project. The aforementioned information-theoretic metrics could be used in an analogous fashion as the Maintainability Index \cite{coleman1994metrics}, where Halmstead volume \cite{halstead1977elements}, cyclomatic complexity \cite{mccabe1976complexity} and lines of code (loc) are combined to produce a value that represents the complexity of the source code. They could even be integrated with classic metrics of complexity to result in a more encompassing metric.

In the following sections, we illustrate our vision for how information-theoretic metrics could be integrated into graphical user interfaces that developers and maintainers use to manage their workflow.

\subsection{Providing Feedback at Development Time}

Feedback during development time is crucial for maintaining the quality, consistency, and overall health of a software project. Real-time feedback enables developers to detect and correct issues as they arise, rather than discovering them later in the development cycle, when they may be more difficult and costly to address \cite{boehm1995cost, boehm2009software}. This proactive approach to software development ensures that problems such as code complexity, deviations from coding standards, or the introduction of bugs and/or unnecessary complexity are identified and resolved quickly. Note, though, that we are making no claims regarding the equivalence between surprising and defective code; here we focus on surprisal as a potential source of complexity, which may in turn be source of defects \cite{McConnell:2004:CCS:1096143, mcconnel2018techdebt}.

Furthermore, real-time feedback fosters a culture of continuous improvement within development teams. It encourages developers to adhere to best practices as they are constantly aware of the quality and complexity of their code. This ongoing feedback loop not only improves the individual developer's skills but also contributes to the overall robustness and maintainability of the software project. By integrating automated detection systems that alert developers to surprising events and potential problems as they code, development teams can maintain high standards of code quality and reduce the long-term costs associated with software maintenance.

\subsection{Examples of GUI-based Warnings and Suggestions}

Incorporating feedback mechanisms directly into the development environment can significantly enhance the developer's workflow. For instance, by integrating entropy-based anomaly detection into the coding process, developers receive immediate notifications of unusual changes in code complexity. This allows them to assess whether these changes are intentional and beneficial or if they could introduce unintended consequences. Early detection and correction of such problems prevent the accumulation of technical debt and reduce the likelihood of future maintenance challenges.

This section provides illustrations of how the hypothetical surprisal Index described in Section \ref{vision_automated_detection} may assist these two kinds of stakeholders in managing source code complexity as early as possible during the development life cycle.

\subsubsection{Development Time - Integrated Development Environment (IDE)}

Integrated Development Environments (IDEs) have become indispensable tools for developers, offering a comprehensive suite of features that significantly enhance productivity and code quality. One of the most critical aspects of modern IDEs is their powerful static analysis capabilities. These tools analyse code without executing it, allowing developers to identify potential errors, security vulnerabilities, and code smells early in the development process. By providing real-time feedback, static analysis helps developers adhere to coding standards, reduce the likelihood of introducing bugs, and maintain a high level of code quality.

Beyond static analysis, IDEs offer a range of functionalities that assist developers in their day-to-day tasks. Features such as intelligent code completion, syntax highlighting, and refactoring tools streamline the coding process, enabling developers to write code more efficiently and with fewer errors. IDEs also typically include integrated debugging tools that allow developers to step through their code, inspect variables, and diagnose issues with greater precision.

Moreover, IDEs often come with built-in support for version control systems, making it easier for developers to manage their code repositories, track changes, and collaborate with others. This integration simplifies workflows and ensures that teams can work together seamlessly, even in complex projects.

\begin{figure}
    \centering
    \captionsetup{justification=centering}
    \includegraphics[width=\textwidth]{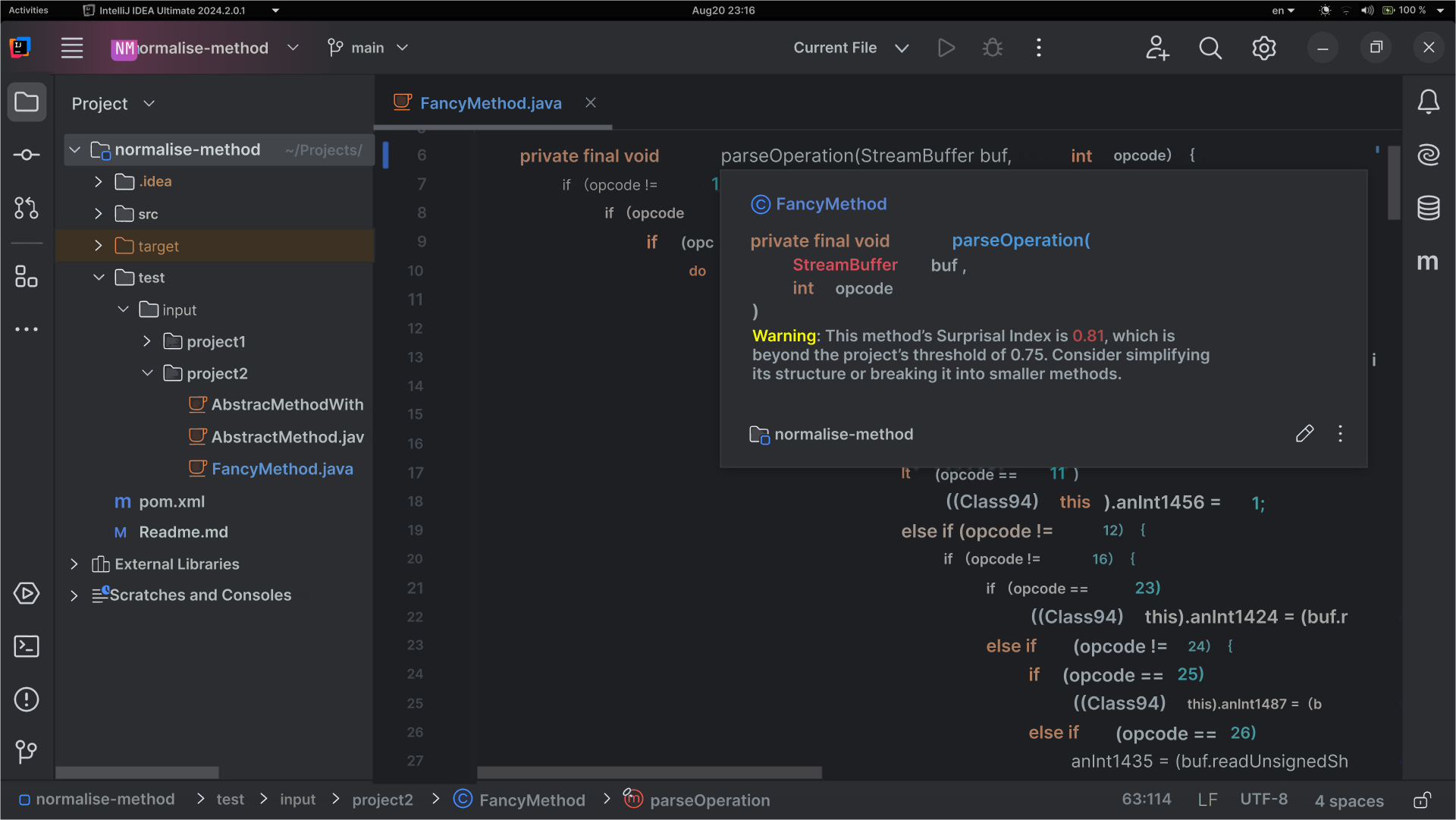}
    \caption{IDE-generated message when surprisal Index is beyond threshold}
    \label{fig:ide_high_surprisal}
\end{figure}

Figure \ref{fig:ide_high_surprisal} shows an example of an IDE message where the complexity of a method, as measured by its Surprisal Index, goes beyond that which would be accepted by a configurable parameter in the project's settings. Similarly to linters that highlight code snippets based on the structure of the Abstract Syntax Tree (AST) or the Control Flow Graph (CFG) of the program, the static surprisal analyser makes a recommendation to the developer before the code is committed.

\begin{figure}
    \centering
    \captionsetup{justification=centering}
    \includegraphics[width=\textwidth]{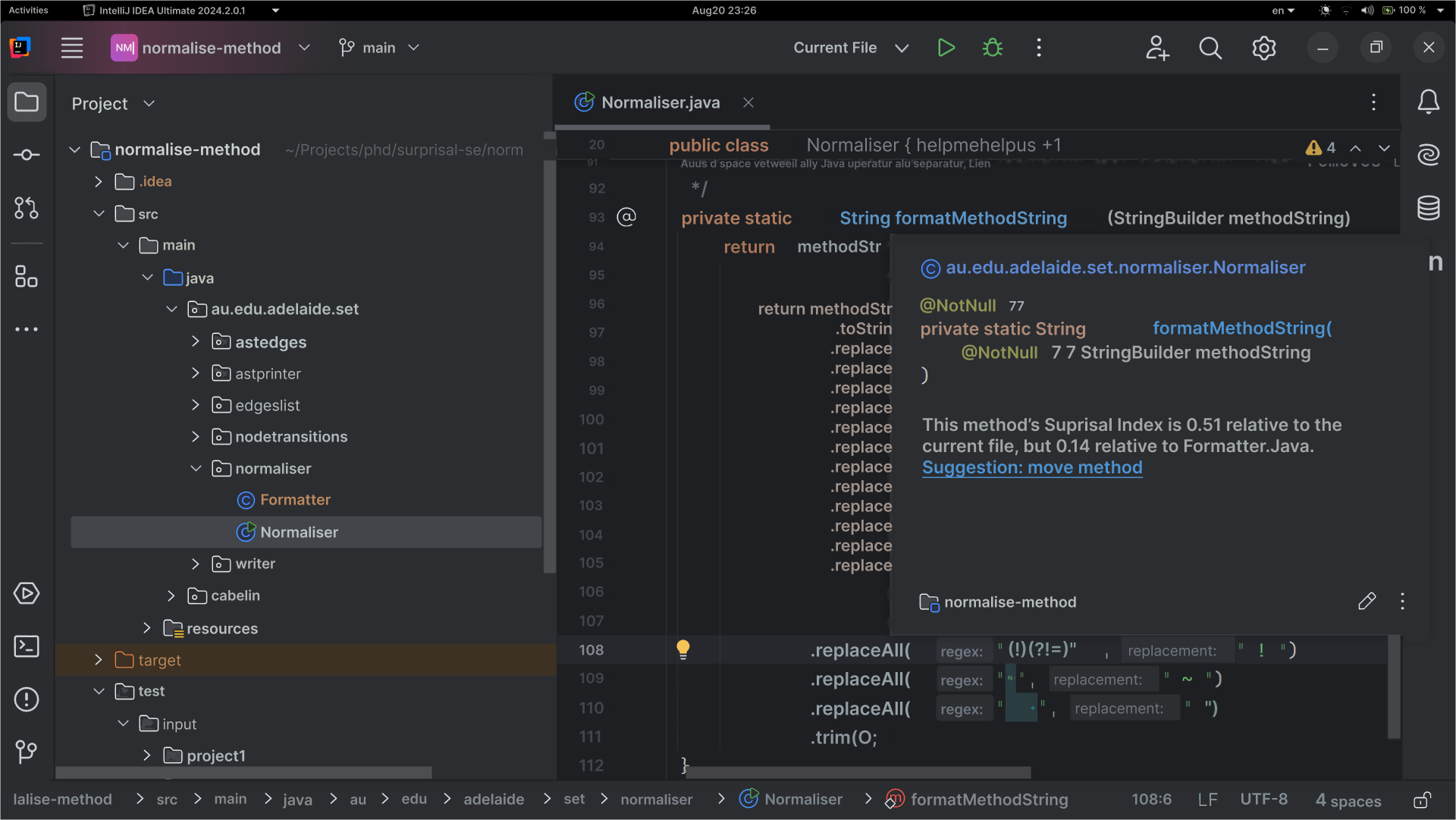}
    \caption{IDE-generated message when surprisal Index is beyond threshold}
    \label{fig:ide_surprisal_suggestion}
\end{figure}

Figure \ref{fig:ide_surprisal_suggestion} depicts a situation where we can leverage information theory's ability to measure contextual similarity, therefore producing a suggestion to move a method over to a class that is ``less surprised'' by the method being introduced.

\subsubsection{Merge Request Time - Version Control System (VCS)}

Maintainers who will assess propose changes submitted by developers need to make judgement calls regarding where to direct their attention. Providing them with a quantitative measure of the suspected impact of a change in the complexity of the code base may not only help them direct the attention to more pressing issues, but also to be more critical when assessing them.

\begin{figure}
    \centering
    \captionsetup{justification=centering}
    \includegraphics[width=\textwidth]{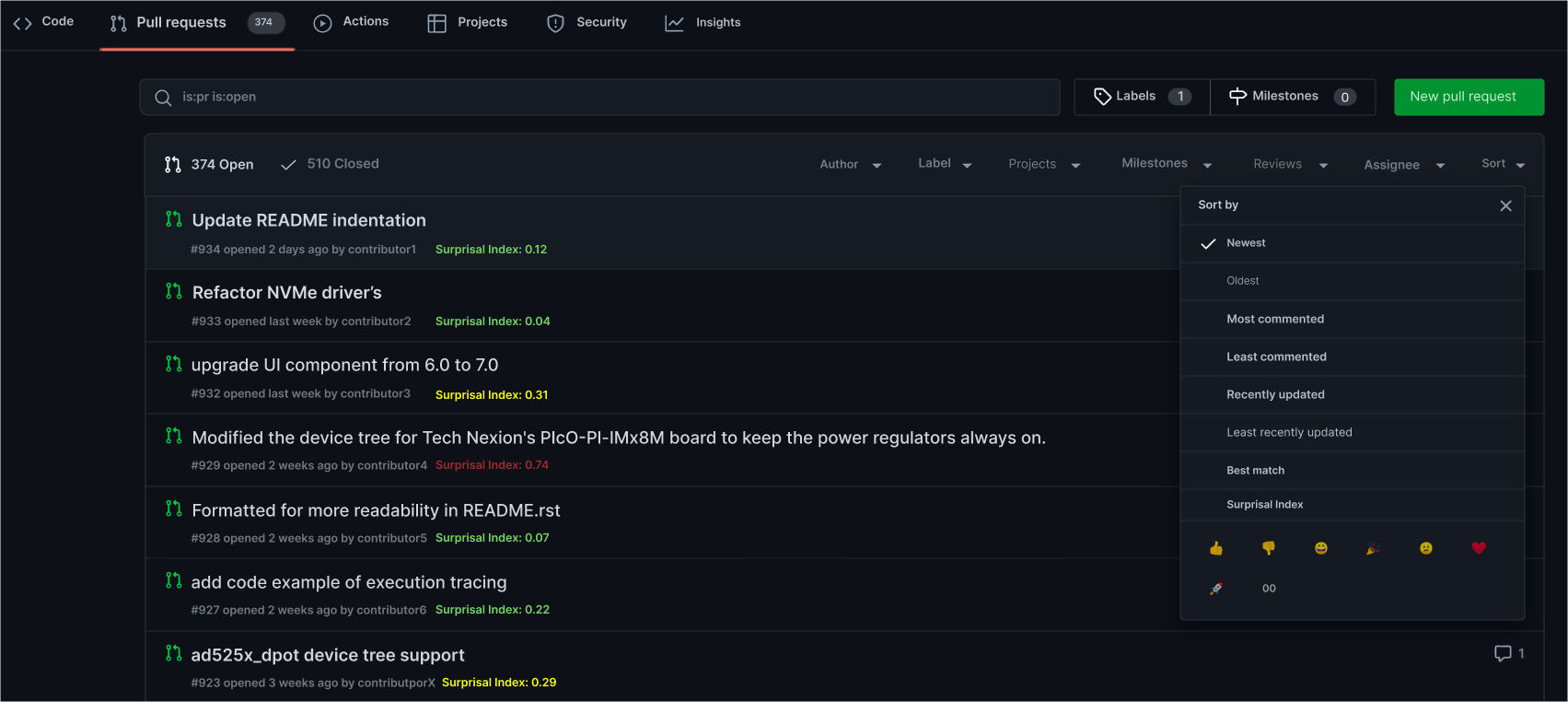}
    \caption{Attaching surprisal Index Information to Merge Requests}
    \label{fig:pr_surprisal_index}
\end{figure}

Figure \ref{fig:pr_surprisal_index} contains an adapted example of the User Interface (UI) of a version control system where the maintainer can immediately see the surprisal Index of each merge request that has been submitted, as well as filter and sort them accordingly. This may help guide several of the maintainer's decisions, such as which ones to analyse first, which ones to delegate, and which ones to include in the next release.

\begin{figure}
    \centering
    \captionsetup{justification=centering}
    \includegraphics[width=\textwidth]{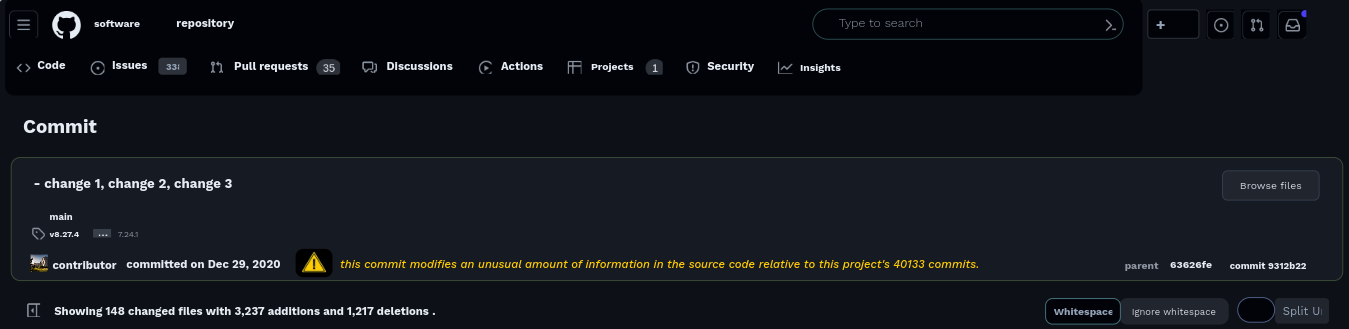}
    \caption{Expanded View of Suprising Merge Request}
    \label{fig:surprising_pr}
\end{figure}

Figure \ref{fig:surprising_pr} illustrates the view of a merge request that was flagged as unusual. The expanded view of each file may in turn provide the Surprisal Index of each file before and after the change, which can further guide the reviewer's attention to the files that were more severely impacted by the change.

\subsection{Discussion}

The integration of automated detection systems based on information theory into the software development life cycle can be a good complement to traditional static analysis tools to enhance the efficiency and effectiveness of software construction. By focusing on unusual changes in code complexity and other entropy-based anomalies, these tools can help developers and maintainers manage the increasing complexity of modern software projects with greater ease. We acknowledge once again the speculative nature of this discussion and propose ways to validate or refute them in \autoref{future_work}, where we we describe future research opportunities.

One of the primary benefits of these systems is the reduction of cognitive load for developers and maintainers. In large and evolving codebases, the sheer volume of changes can make it difficult to identify those that may have far-reaching implications. Automated detection tools that highlight surprising or significant changes enable developers to focus their attention on areas that require immediate attention, thereby addressing them before they become too expensive. By catching potential issues early in the development process, developers can address them before they become deeply embedded in the codebase, reducing the need for costly and time-consuming refactoring efforts later on. This proactive approach to software development not only saves time but also improves the overall quality and reliability of the software, as fewer issues are likely to reach production. This targeted approach not only improves the quality of the code, but also reduces the time and effort required to manage the project's growth.

For maintainers, the ability to quickly identify and assess high-impact changes is crucial for maintaining the long-term health of the project. Tools that quantify the complexity and potential risks associated with changes, such as the proposed Surprisal Index, provide maintainers with actionable insights that can guide their decision-making processes. By prioritising the review of changes that are likely to introduce significant complexity or deviate from established patterns, maintainers can ensure that their time and attention are being directed towards important issues.

The discussion around integrating these tools into Integrated Development Environments (IDEs) and Version Control Systems (VCS) highlights their potential to make developers' lives easier. By providing real-time feedback during development and detailed analysis during code reviews, these tools can help developers adhere to best practices, avoid introducing unnecessary complexity, and maintain a high standard of code quality. The ability to receive immediate warnings and suggestions within the IDE, as well as to prioritise merge requests based on their potential impact, streamlines the workflow, and allows developers and maintainers to work more efficiently.

\section{Comparison With ChatGPT}
Over the past few years, large language models (LLMs) have seen a surge in capability and usage, with applications ranging from education \citep{kasneci2023chatgpt} to quantitative reasoning \citep{lewkowycz2022solving}. Given that software written in high-level programming languages tends to exhibit a high degree of "naturalness" \citep{hindle2016naturalness} and that our definition of textual entropy is based on words extracted from source code (which correlate highly with our definition of structural entropy), it is reasonable to contrast our approach against models like ChatGPT-4 \citep{openai2024gpt4technicalreport} and highlight their relative strengths and weaknesses across several dimensions.

In this section, we provide a high-level comparison between our approach and ChatGPT-4, focusing on fundamental differences rather than an exhaustive evaluation. Our intent is to highlight key conceptual and methodological distinctions, rather than to conduct a detailed benchmark. A comprehensive comparison involving multiple LLMs, including ChatGPT, DeepSeek, Claude, and Gemini, is deferred to future work.

\subsection{Method and Scope of Analysis}
ChatGPT excels in its adaptability and breadth of application. It can interpret a wide variety of inputs, such as natural language commit messages, pull request discussions, or user feedback, offering contextual insights that are outside the scope of entropy-based analysis. While our approach is limited to quantifying code changes, ChatGPT can generate explanations, summarize repository histories, or highlight potential issues based on patterns in the data, including non-code artifacts, which therefore may allow it to better encode expectations and surprisal.

Our method provides a deterministic and granular analysis of software evolution by quantifying entropy changes commit by commit. This level of granularity allows for precise (quantitative) tracking of textual and structural entropy metrics, enabling the detection of patterns and trends that are otherwise difficult to capture. Unlike ChatGPT, our approach does not rely on training a model with hundreds of billion parameters \citep{llmfewshot2020} or generalizing across datasets. It directly analyses the raw data from repositories, making it inherently reproducible and free from probabilistic biases. Moreover, our approach allows us to input and measure the entire code base at each commit.

\subsection{Explainability and Control} \label{explainability}
Our system is fully transparent and lightweight, relying on straightforward calculations derived from token and AST histograms. This makes it highly explainable and easy to integrate into workflows. Developers retain full control over their data and processes, as all computations are performed locally without reliance on external APIs or cloud services. This ensures privacy and security, particularly for sensitive or proprietary projects. Submitting an entire code base at each commit to the ChatGPT API could prove both unsafe and impractical relative to integrating our approach to deployment pipelines that run code quality checks.

ChatGPT operates essentially as a black-box model, which can be a limitation for those requiring deterministic and explainable results. However, its flexibility compensates for this by providing high-level, partial interpretability through natural language explanations. It is particularly effective in translating technical complexities into plain language, making it an invaluable tool for cross-disciplinary teams or for educating less technical stakeholders. Still, its non-deterministic nature is known to produce different results to the same prompt, and researchers have exposed limitations in its mathematical reasoning \citep{anonymous2024gsmsymbolic}. Our approach relies on mathematical measurements who may differ between two commits in the order of thousandths of bits, which LLMs may struggle to capture qualitatively.

\subsection{Resource Efficiency}
Our method is highly efficient in terms of computational resources. The tools used for data collection and analysis amount to less than 200kB, and even large project datasets do not exceed 60MB. This makes it feasible to integrate into CI/CD pipelines or other resource-constrained environments.

ChatGPT’s resource requirements are significantly higher, both in terms of model size, training time and computational power. It is true that, once deployed, it may eliminate the need for specialized tools by serving as a general-purpose assistant capable of handling a wide range of tasks, which amortises its cost whilst providing broad applicability. However, as discussed in Section \ref{explainability}, it may be impractical and costly to employ it to analyse all the code at each commit. 

\subsection{Contextual Adaptability}
Our system’s focus on entropy-based measurements ensures consistency and precision, but it is limited to analyzing code and its immediate structure. It does not require additional prompts or user guidance, offering a streamlined, deterministic output.

ChatGPT’s adaptability is its core strength. It can handle diverse scenarios, from summarizing complex code to analyzing project metadata, thus enabling to form a large context against which it will judge how surprising a change may be. This may provide considerable advantage over our fixed-methodology that does not consider other sources of information generated during software development that may be just as important as the code for defining expectations about its evolution.

\subsection{Anomaly Detection and Insights}
While ChatGPT does not natively perform anomaly detection, it can infer patterns and provide qualitative insights into repository activity when prompted. It is particularly useful for reasoning about the implications of changes, such as whether a spike in complexity aligns with project goals or user requirements. This ability to contextualize changes complements the quantitative outputs of our approach.

Our entropy-based anomaly detection provides a systematic method for identifying unusual events, such as spikes in complexity or architectural changes. By processing every commit and tracking entropy changes, our system highlights potentially significant modifications that may introduce technical debt or increase maintenance costs. The categorization of these events allows maintainers to proactively address issues, a capability not inherent in ChatGPT.

\subsection{Summary}
Our approach excels in precision, explainability, and efficiency, making it ideal for deterministic measurements of software evolution and anomaly detection. ChatGPT, with its flexibility and contextual reasoning, is better suited for tasks requiring interpretation, natural language understanding, and adaptability. From a practical standpoint, integrating our approach with a pre-trained LLM that can be used within the project owners' development and deployment environments may give rise to a comprehensive analysis framework, leveraging the strengths of both deterministic metrics and probabilistic inference.

\section{Threats to validity} \label{limitations}

Ensuring the validity of research findings is crucial for drawing accurate conclusions. In this section, we discuss internal, external, and construct threats to the validity of our findings.

\textbf{Internal validity} represents our concerns about the extent to which it is possible to rule out alternative explanations for our findings~\cite{brewer2000research}. Although we have observed that the correlation between entropy and certain other source code complexity metrics (e.g., line/method/token counts) was weak to moderate across the vast majority of the projects we analysed, we do not know the extent if our information-theoretic methods would be as valuable to practitioners as static analysers that measure cyclomatic complexity, coupling or cohesion. The fact that the correlations are low doesn't mean that entropy might correlate with the abstract concept(s) that complexity metrics are supposed to measure.

The foremost internal threat in this work is our reliance on \textbf{manual inspection and subjectivity} when labelling and unusual event. The manual analysis of detected events introduces potential bias and subjectivity, particularly in categorising events as true or false positives. Despite efforts to validate the labelling process, the subjective nature of this task could lead to inconsistencies in how events are evaluated, potentially affecting the study's internal validity. As we are not involved in the development of the repositories we analysed, we have no domain knowledge or expertise about their respective domains. Moreover, we are not aware of all the communication that developers and maintainers exchange, which may in turn affect their expectations about certain changes; for example, there may be changes that introduce (or remove) a lot of information to (or from) the project, but that were nonetheless agreed to and expected, and therefore not surprising at all. Validating the commits we identified as surprising would entail submitting the events to each maintainer of the projects we analysed and asking them to label each event according to their knowledge of the project in a similar manner as was done in previous research regarding unusual events \cite{TREUDE2018237Unusual}, but this time with a focus on entropy variations. The design and actual conduction of such experiment is part of what we propose as extension to this research.
Finally, file-level entropy metrics may lead our detection to capture changes in total information content that are inflated, thus reducing the precision of our measurements.

\textbf{External threats} are concerned with how our findings generalise to other contexts~\cite{baltes2022sampling}. Our study's findings focus on open-source Java projects, which might limit the generalizability of our results in regards to structural entropy. Textual entropy may also be affected by the domain of the systems under construction, as well as by conventions established by either the community of the respective programming language or by the maintainers themselves. More broadly, one of the main characteristics of open-source are voluntary contributions under no strict management or timelines, which may limit the extent to which our findings apply to proprietary software development. We have also not considered \textit{interaction effects}, as entropy as a measure of software evolution might interact with other project characteristics, such as the size of the project team, the frequency of commits, or the nature of the software (e.g., web application vs. embedded system), or the philosophy of development (e.g. Agile vs Waterfall). These interactions could limit the extent to which the findings can be generalised to different types of software projects or development practices.

Since software engineering is primarily a human activity, factors like domain expertise and agreements between stakeholders may greatly impact whether an event is considered unusual or not. Ultimately, the effectiveness of our approach may vary depending on the type of project (e.g., small vs. large-scale projects, different application domains). Unless we integrate stakeholders' opinions when labelling an event as unusual, the results might not be fully applicable to projects with different characteristics, limiting the external validity. 

\textbf{Construct validity threats} refer to the degree to which our definitions and methods faithfully measure the intended properties~\cite{external_validity}. Our calculation of the total information content of a project at any point in time consists in calculating the entropy of each of the changed files in a commit separately. Due to entropy being closely related to the frequencies of the tokens/AST edges, our measurement of complexity at each point is likely inflated; two files that are belong to the same unit of abstraction, like modules, packages and namespaces, are likely to share common functionality, structure and language tokens. Therefore, considering the related files in isolation does not account for the aggregate frequency of tokens and edges with high precision. Not only can this cause inflation in our metrics, but it may render historical entropy graphs whose rate of growth is smaller than the ones we obtained. Moreover, considering wider scopes of context may produce different correlation values between entropy and classic metrics. Considering wider contexts - such as namespaces, modules and packages - entails considerable engineering effort and is scope of future work.

The reliance on entropy as the sole metric for detecting unusual events could limit the validity of our constructs. \textbf{Single method bias}: If entropy alone does not adequately capture all dimensions of software complexity or code changes, other important factors might be overlooked. A more comprehensive approach incorporating multiple metrics could provide a fuller understanding of software evolution. Although we used rigorous statistical definitions for entropy and for detecting outliers, a definition of an unusual or unexpected event that does not encode the agreements and knowledge of the project owners is bound to be incomplete. 
Another threat involves the use of Krippendorff's alpha to measure agreement of the separate labelling of events by the first two authors, due to issues of reliability estimates with the statistics itself \cite{THOMPSON1988949}. Moreover, none of the authors had contextual understanding of the projects being analysed. We attempted to mitigate both effects by being conservative when labelling points, i.e., by being willing to accept higher false positive rates, and by selecting the definition of statistical outlier (i.e., z-score) for which we observed the best signal-to-noise ratio.
Since we had to rely on our own judgement when labelling unusual events, the precision of our detection may be vastly inferior to that obtained if events were labelled by the actual maintainers of the code at hand or by applying other forms of automation that would reduce the bias introduced by our subjective assessments. We tried to mitigate this by revising all points twice during analysis, as well as by manually evaluating three separate significant samples of events with $95\%$ confidence interval, as well as by having the author and one of the supervisors assess events independently and obtaining a high level of agreement.

\textbf{On defining an usual event}: Determining whether a commit is surprising is not a trivial task, as it involves understanding the project context, the habits and conventions of its developers and maintainers, and domain knowledge. This can be an internal threat since it affects the precision metrics of the detection of unusual events, whose labelling process involved deciding whether a change was relevant. Since surprisal is ultimately dependent on context and in maintainer's expectations, it becomes difficult to extend our findings from one project to another.

\section{Future work} \label{future_work}
As mentioned in \autoref{limitations}, considering wider scopes over which entropy is calculated may provide a more accurate picture of the actual information content of the project as a whole. This could be achieved by applying heuristics to identify the logical boundaries to which groups of files belong (e.g. packages and modules) and consider these wider scopes as the context for entropy calculations. This could produce less inflated metrics, as concepts that are shared amongst files will not be computed in isolation. Including other information-theoretic measures of statistical similarity in the definition of the hypothetical Surprisal Index introduced in \autoref{vision_automated_detection} can also complement the engineering efforts mentioned above to increase the precision of the detection.

Values toward the higher end of the range may mean that a proposed change presents a high increase in complexity, or that there is low contextual similarity between the change and the scope against which the surprisal Index is being measured.

We also discussed in \autoref{limitations} that, as outside observers just collecting data, our opinions on how surprising a commit is are limited and may be inaccurate. Ultimately, developers and maintainers of the project have the final say on whether a change is expected or not. Given that there is evidence that developers are interested in methods and tooling to assist them in identifying unusual events \cite{TREUDE2018237Unusual}, we believe that submitting the events we detected and asking them to label the commits according to their knowledge of the history of the project can help us measure the usefulness of our approach and can also help us identify areas of improvement. In such an experiment, we could also investigate whether automating the detection of surprising or unusual events can reduce the cognitive load maintainers experience as they process the multiple sources of information when making decisions about the evolution of the project.

Extending the analysis to other intermediate representations like Control Flow Graphs (CFG), Lossless Semantic Trees (LST), LLVM/MLIR representations, or other partial intermediate representations may also provide further definitions and insights into structural entropy. These structures encode the semantics of programs, and their analysis may form the basis of a measure of how much information is necessary to encode the meaning of different program constructs (or of programs as a whole), therefore producing a basis of comparison for semantic complexity.

All of these intermediate representations are hierarchical structures, and some patterns seem to naturally increase the complexity of a code base. For example, in an AST, a series of nested method call expressions could be an indicator that the subtree that represents that portion of code yields higher complexity. Moreover, method and/or function declaration nodes whose list of parameters is longer tend to correspond to more complex methods. In such cases, being able to assign weights to nodes with certain properties and/or patterns may produce entropy measures that better represent their perceived complexity.

Ultimately, the development of methods that can assist practitioners in their daily work must consider their needs and provide value \citep{Christakis2016DevelopersNeeds}. We envision the prototyping of a tool that automates the detection, classification, and ranking of unusual events, which could then be offered to maintainers for evaluation and feedback.

\section{Conclusion}
We conducted a statistical assessment of the evolution of the information conveyed by source code during each point of its evolution in 95 open source projects. Our study revealed that projects usually undergo two distinct phases: first, a steep pace of change during its initial stages, followed by a period where the rate of change of information content stabilises but still keeps increasing. While this pattern is consistent for all repositories, the projects differ in the way the information is spread throughout the source code files, with some projects tolerating higher density of information per file, whereas others try to minimise it. In general, while two projects show declining trends in their latest 20\% commits, linear regression indicates a general increase in total information across all projects. However, analysis on a per-file basis reveals varying patterns: strict control in one project, stabilisation after initial chaos in 73 projects, and continuous growth in 21 projects.

We also analysed the correlations between our definitions of source code entropy and classic metrics of source code complexity, such as lines of code, number of tokens and number lines changed, and cyclomatic complexity. We learned that they are weak for the most part, suggesting that entropy may capture dimensions of complexity that are not measured by classic code complexity definitions. 
The correlation between structural and textual entropy varies across projects: 2 show weak correlations, 61 moderate, and 32 strong. Entropy measurements have low to moderate correlation with insertions, deletions, nloc, tokens, and changed methods. The correlation with McCabe’s cyclomatic complexity is notably low, ranging between -0.05 and 0.32. While entropy and insertions-deletions correlation is strongest on average, it varies significantly within projects. These findings imply that code complexity can potentially be measured through entropy of source code representations.

Furthermore, we used our statistical definitions of the entropy of the source code to empirically evaluate Lehman's laws of software evolution~\citep{Lehman1980Laws}, \citep{LEHMAN1984ProgramEvolution}. Our measurements agree with two laws that relate to evolution of program complexity: the law of continuing change and the law of increasing complexity. However, in regards to managerial aspects of the software development process, our assessment disagrees with laws that are concerned with ensuring smooth evolution of the projects. We believe that this is mostly due to differences in the context of the projects, since open source projects have looser management practices than their private counterparts.

We concluded our work by applying anomaly detection to the distribution of variation of entropy as a means to capture unusual changes and investigated whether such a way of capturing events can assist maintainers in keeping on par with important changes to the source code. Almost two-thirds of events concern internal system features, with rare occurrences impacting user-facing complexity. Events affecting users often involve significant file changes, which is crucial for maintainers. Refactors and bug fixes are minimal (3.75\%), while removal of old/unused code dominates false positives (31\%). Limiting event detection window increases the number of detected events by 20\%, but maintaining strict criteria for false positives decreases precision. Criteria stringency depends on project history, affecting the practicality of automatic entropy-based event detection.

During this endeavour, we learnt that while we applied strict statistic methods for defining and labelling outliers, the relevance/impact of each change is rather context-dependent, since it involves maintainers' conventions and expectations, which ultimately means that an entropy-based system for detecting unusual code changes may need to incorporate stakeholders' assessments of the events we capture in order for such a system to become more precise and useful.  

\section{Data Availability Statement}
All of the source code for data collection, parsing, processing and analysis are available in the supplementary material \citep{supp_material}, as well as manual point analysis, visualizations and charts and inter-rater agreement measurements. For further explanation of how the material is organised, please refer to its \texttt{README.me} file in the root folder of the data repository.

\section{Acknowledgements}
The authors thank the 2nd International Workshop on NL-based Software Engineering (NLBSE 2023) for reviewing the initial manuscript~\citep{nlbse2023entropy} and for suggesting an assessment of our data against Lehmanh's laws of software evolution~\citep{Lehman1980Laws}. We also thank Fabian Beck for suggesting improvements for our visualisation of correlations between entropy and classic complexity metrics.

\section{Declarations}

\paragraph{Compliance with Ethical Standards:} The authors confirm that this work complies with ethical standards.

\paragraph{Conflict of Interest/Competing Interests:} The authors declare no competing interests. However, Sebastian Baltes and Christoph Treude are members of the EMSE Editorial Board, which is disclosed here for transparency.

\paragraph{Funding:} No funding was received for conducting this study.

\paragraph{Ethical Approval:} No research involving humans or animals was conducted in this study. All code bases analyzed in this work were publicly available as open-source projects at the time of analysis.

\paragraph{Informed Consent:} No research involving humans or animals was conducted in this study. Therefore, no informed consent was required.

\subsection*{Author Contributions}
\begin{itemize}
    \item \textbf{Adriano Torres}: Conducting the empirical study, including data collection and analysis. Literature review. Writing the empirical assessment section and summarizing the results (e.g., patterns of entropy evolution). Developing and documenting the software tools used for analysis. 
    \item \textbf{Sebastian Baltes}: Conceptualization and formulation of research questions. Overseeing statistical evaluations and final data validations. Experiment design.
    \item \textbf{Christoph Treude}: Reviewing and contributing to the framing of research questions and results. Reviewing methods and contributions. Providing related work references.
    \item \textbf{Markus Wagner}: Advising on statistical techniques used in the study and literature review methods. Providing oversight on the broader scientific narrative and its positioning in the field.
\end{itemize}

\bibliographystyle{spbasic}
\bibliography{sn-bibliography}
\end{document}